\documentclass{elsarticle}

\usepackage{bm}
\usepackage{lineno,hyperref}
\usepackage{graphicx}
\usepackage{color}
\usepackage{tabularx}
\usepackage{amsmath,amssymb}

\modulolinenumbers[5]

\journal{Computer Physics Communications}

\graphicspath{
	{./}
	{./figures/}
	{./BG_results/}
	{./BG_results/final_complect/}
}

\begin{document}

\begin{frontmatter}

\title{Numerical analytic continuation of Euclidean data}

\author[ECT]{Ralf-Arno~Tripolt\corref{mycorrespondingauthor}}
\cortext[mycorrespondingauthor]{Corresponding author.}
\ead{tripolt@ectstar.eu}

\author[TOK,KEI1,KEI2]{Philipp~Gubler}

\author[BLF]{Maksim~Ulybyshev\fnref{newaddress}}
\fntext[newaddress]{Address after March 2018: Universit\"at W\"urzburg, 97074 W\"urzburg, Germany}

\author[JLU]{Lorenz~von~Smekal}

\address[ECT]{European Centre for Theoretical Studies in Nuclear Physics and related Areas (ECT*) and Fondazione Bruno Kessler, Villa Tambosi, 38123 Villazzano (TN), Italy}

\address[TOK]{Advanced Science Research Center, Japan Atomic Energy Agency,	Tokai, Ibaraki 319-1195, Japan}
\address[KEI1]{Department of Physics, Keio University, Kanagawa 223-8522, Japan}
\address[KEI2]{Research and Education Center for Natural Science, Keio University, Kanagawa 223-8521, Japan}

\address[BLF]{Institut f\"ur Theoretische Physik, Universit\"at Regensburg, 93053 Regensburg, Germany}
        
\address[JLU]{Institut f\"ur Theoretische Physik, Justus-Liebig-Universit\"at, 35392 Giessen, Germany}

\begin{abstract}
In this work we present a direct comparison of three different numerical analytic continuation methods: the Maximum Entropy Method, the Backus-Gilbert method and the Schlessinger point or Resonances Via Pad\'{e} method. First, we perform a benchmark test based on a model spectral function and study the regime of applicability of these methods depending on the number of input points and their statistical error. We then apply these methods to more realistic examples, namely to numerical data on Euclidean propagators obtained from a Functional Renormalization Group calculation, to data from a lattice Quantum Chromodynamics simulation and to data obtained from a tight-binding model for graphene in order to extract the electrical conductivity.
\end{abstract}


\end{frontmatter}


\section{Introduction}
\label{sec:introduction}

The necessity to perform an analytic continuation of numerical data is a common but also ill-posed and therefore difficult problem in physics. The analytic continuation problem in general refers to the task of extending the domain of a function beyond the regime where it is known or where there are data points available. It is encountered for example in Euclidean Quantum Field Theory (QFT) when one aims at reconstructing real-time correlations or spectral functions based on some discrete and finite set of data points along the Euclidean (imaginary) time axis. Euclidean correlation functions in principle contain the complete information of a field theory in thermal equilibrium. The inverse problem of extracting this information from these correlations is often NP-hard, however. While their analytic continuation is unique in principle, with suitable boundary conditions \cite{Baym1961,Landsman1987}, it requires infinitely many data points of infinite precision. The spectral representation of a Euclidean time correlation function at zero temperature, for example, is given as a Laplace transform of the corresponding spectral function. Knowing the correct analytic continuation of this correlation function is equivalent to knowing its spectral function. When dealing with a finite set of data points of finite accuracy, however, the inverse Laplace transform becomes an ill-posed numerical problem.

There are several numerical continuation methods available in the literature that aim at obtaining the best possible reconstruction of spectral functions. For example, the Maximum Entropy Method (MEM) \cite{Jaynes1957,Jaynes1957a,PressTeukolskyVetterlingEtAl2002}, the Backus-Gilbert (BG) method \cite{BackusGilbert1968,G.Backus1970,PressTeukolskyVetterlingEtAl2002}, the Schlessinger Point (SP)
or Resonances Via Pad\'{e} (RVP) method \cite{Schlessinger1966, TripoltHaritanWambachEtAl2016}, or a Tikhonov regularization \cite{Dudal:2013yva}, which allows to probe unphysical (non positive-definite) spectral densities, see also \cite{HobsonLasenby1998}, have been proposed. They all have different strengths and different regimes of applicability. The question as to which of the methods will give the best reconstruction therefore depends on the particular problem to which they are applied.

In this work we present a direct comparison, both in QCD and in condensed matter systems, of three of those numerical analytic continuation methods: MEM, the BG method and the SP (or RVP) method. 
While all three methods have been used and tested in the literature before, see e.g.~\cite{GunnarssonHaverkortSangiovanni2010} for MEM and the SPM in the context of estimating the optical conductivity in a condensed matter system, \cite{Vidberg:1977, landau2015, landau2016, Idan2016, MarkoReinosaSzep2017, LeeLiMonien2008, OsolinZiitko2013} for the SPM as a numerical analytic continuation technique in various situations, \cite{Jarrell:1996,Asakawa:2000tr} for the MEM, and \cite{BrandtBG} for the BG method, the SP method has only rarely been used in a QCD setting and is applied for example to lattice QCD data for the first time in this work to our knowledge.
 
We will focus on reconstructing the spectral function $\rho(\omega)$ based on a finite set of data points for a correlation function $G^E(\tau)$ in Euclidean (imaginary) time  $\tau$. At finite temperature, $T=1/\beta$, the Euclidean correlation function is related to the spectral function by the periodic extension of 
\begin{align}
\label{eq:Lehmann:tau}
G^E(\tau)&=\int_{0}^{\infty}d\omega\rho(\omega)\frac{\cosh(\omega(\tau -\beta/2))}{\sinh(\beta\omega/2)} \, , \quad \mbox{for} \;\; \tau \in [0,\beta]\, .
\end{align}
MEM and the BG method aim at inverting this integral using input points for $G^E(\tau)$, while we will use the SP method to perform an analytic continuation of the Euclidean propagator $D^E(p^0)$, where $p^0$ is the Euclidean (Matsubara) frequency. After the analytic continuation, the retarded propagator $D^R(\omega)$
is defined at real frequencies $\omega = -i p^0 $, which yields the spectral function from 
\begin{align}
\rho(\omega)=-\frac{1}{\pi}\, \text{Im}D^R(\omega).
\end{align}

The discrete data points for $D^E(p^0)$ are obtained from those of $G^E(\tau)$ via a discrete Fourier transform in Secs.~\ref{sec:model_study}, \ref{sec:Lattice} and \ref{sec:conductivity} where we study data for the periodic Euclidean correlation function $G^E(\tau)$ with period $\beta = 1/T$ as in Eq.~(\ref{eq:Lehmann:tau}) at the $N$ discrete times $\tau_m = \beta m/N$. In order to reconstruct the corresponding spectral function $\rho(\omega)$ with the SP method, we will first use the discrete Fourier transform,
\begin{align}
\label{eq:FT}
D^E_N(p^0_n)&=\frac{\beta}{N}\sum_{m=0}^{N-1} G^E\left(\tau_m\right) \, \exp\left(i 2\pi  nm/N\right),
\end{align}
to obtain $D^E_N(p^0_n)$ as an approximation to the Euclidean propagator $D^E(p^0_n)$ obtained in the limit ${N\rightarrow\infty}$ as the Fourier series coefficients of the continuous periodic $G^E(\tau)$. The error induced by using a finite number of points in the Fourier transform will make it more difficult to obtain a reliable reconstruction with the SP method if the number of input points is too small, see Sec.~\ref{sec:model_study}.

In Sec.~\ref{sec:FRG}, the inverse Fourier transform will be used in order to obtain $G^E(\tau)$, as needed by MEM and the BG method, from the Euclidean propagator $D^E(p^0)$ as given by the FRG calculation at a finite set of discrete Matsubara frequencies $p^0_n = 2\pi Tn $. This transform is given by
\begin{align} \label{eq:ZNFT}
G^E_N(\tau_m)&=T\sum_{n=-N/2+1}^{N/2}D^E\left(2\pi T n\right)\, \exp\left(-i 2\pi  n m/N \right),
\end{align}
where $G^E_N(\tau_m)$ is an approximation to the full Euclidean correlation function $G^E(\tau)$ at the discrete Euclidean times $\tau_m = \beta m/N$ that is exact only in the limit ${N\rightarrow\infty}$. We checked that the error introduced by the discrete Fourier transform is small in Sec.~\ref{sec:FRG} where we use $N=2048$ and that it does not affect the quality of the reconstruction.

In Sec.~\ref{sec:methods} we will discuss the three different analytic continuation methods used in this work. They are then applied to study a simple model for a spectral function in Sec.~\ref{sec:model_study} where we will also study their different regimes of applicability depending on the number of input points for $G^E(\tau)$ and the statistical error of those points. We then apply the three methods to three different situations that are typically encountered in field-theory applications: first, in Sec.~\ref{sec:FRG}, to data obtained from a numerical calculation employing the Functional Renormalization Group (FRG) \cite{Tripolt2014}, then, in Sec.~\ref{sec:Lattice}, to the lattice QCD data for the vector-meson correlation function from Ref.~\cite{BrandtFrancisJaegerEtAl2016}, and finally,  in Sec.~\ref{sec:conductivity}, to data for the electromagnetic current correlator obtained from a tight-binding model for graphene in order to extract the electrical conductivity \cite{conductivity1}.

\section{Analytic continuation methods}
\label{sec:methods}

\subsection{Maximum Entropy Method}
\label{sec:MEM}

The basic idea of the Maximum Entropy Method is to make use of all the
available information about the spectral function (including properties such as
positive definiteness and asymptotic values) and to choose it to have a large 
entropy (for a precise definition, see below) while at the same time satisfying the 
conditions given for instance by Euclidean time data.
More specifically, the method relies on Bayes' theorem of probability theory, 
\begin{equation}
P[\rho|GI] = \frac{P[G|\rho I] P[\rho|I]}{P[G|I]}, 
\label{eq:bayes}
\end{equation} 
where $P[\rho|GI]$ stands for the conditional probability of the spectral function $\rho$ to have a 
certain form, given the information on the Euclidean time data $G$ and other prior information $I$ about the spectral function. 
The goal is then to find the specific form of $\rho$ that maximizes $P[\rho|GI]$. On the right hand side of 
Eq.\,(\ref{eq:bayes}), there are two factors that depend on $\rho$, which both should be taken into 
account. The first one, $P[G|\rho I]$, is the so-called likelihood function and can be given as 
\begin{equation}
\begin{split}
P[G|\rho I] &= e^{-L[\rho]}, \\
L[\rho] &= \frac{1}{2}
\sum_{i,j}
\bigl[G^{E}(\tau_i) - G^{\rho}(\tau_i) \bigr] C^{-1}_{ij} \bigl[G^{E}(\tau_j) - G^{\rho}(\tau_j) \bigr] \\ 
&= \frac{1}{2} \sum_{i} \frac{\bigl[G^{E}(\tau_i) - G^{\rho}(\tau_i) \bigr]^2}{\eta^2(\tau_i)},
\end{split}
\label{eq:likelihood}
\end{equation}
where $C_{ij}$ is the covariance matrix and 
$\eta(\tau_i)$ stands for the error of the discrete Euclidean time data $G^{E}(\tau_i)$ at $\tau_i$. 
In going from the second to the third line in Eq.\,(\ref{eq:likelihood}) we have ignored correlations 
between different $\tau$ values, which generally is not possible for real lattice QCD data. 
The dependence on $\rho$ comes from $G^{\rho}(\tau)$, which is the integral on the right hand side of Eq.\,(\ref{eq:Lehmann:tau}). 
As it is clear from the above definition, maximizing $P[G|\rho I]$ is equivalent to ordinary $\chi^2$-fitting. If however the number of 
input data points on the Euclidean time axis is not large enough (which is usually the case), maximizing $P[G|\rho I]$ will not lead to 
a unique solution for $\rho$. 
The second term in Eq.\,(\ref{eq:bayes}), $P[\rho|I]$, the ``prior probability"
therefore plays an important role in this respect by selecting 
among the infinite number of possible solutions the one most 
closely resembling the prior information.
It is given as 
\begin{equation}
\begin{split}
P[\rho|I] &= e^{\alpha S[\rho]}, \\
S[\rho] &= \int_0^{\infty} d\omega \Bigr[ \rho(\omega) - m(\omega) - \rho(\omega)\log \Bigl( \frac{\rho(\omega)}{m(\omega)} \Bigr) \Bigl], 
\end{split}
\label{eq:prior}
\end{equation}
where $S[\rho]$ is known as the Shannon-Jaynes entropy and 
the function $m(\omega)$ is called the ``default model''. One can show that $P[\rho|I]$ becomes largest when $\rho(\omega) = m(\omega)$. In this work we will always use a constant as the default model.

Assembling the above terms, we get the probability $P[\rho|G I]$ as 
\begin{equation}
\begin{split}
P[\rho|G I] &\propto  P[G|\rho I] P[\rho|I]  \\
&= e^{Q[\rho]}, \\
Q[\rho] & \equiv \alpha S[\rho] - L[\rho]. 
\end{split}
\label{eq:finalprob}
\end{equation}
To find a $\rho(\omega)$ for which $P[\rho|G I]$ is largest, one hence has to maximize $Q[\rho]$. 
One can in fact show that the maximum of $P[\rho|G I]$ is unique if it exists \cite{AsakawaNakaharaHatsuda2000}. 
The real and positive parameter $\alpha$ appearing in Eq.\,(\ref{eq:finalprob}) is however not yet determined. The most common way to handle it (which we will follow in this work), is to define a conditional probability of $\alpha$ to take a certain value and to then average $\rho_{\alpha}(\omega)$, which gives a maximum $Q[\rho]$ for a fixed $\alpha$, over this probability. For the details of this procedure, we refer the reader to \cite{AsakawaNakaharaHatsuda2000} and the 
references cited therein. Furthermore, for the numerical task of maximizing $Q[\rho]$, we use Bryan's algorithm \cite{Bryan1990}, 
which is described in some detail in \cite{Jarrell:1996,AsakawaNakaharaHatsuda2000}. 

Looking at Eqs.\,(\ref{eq:likelihood})--(\ref{eq:finalprob}), one can consider two limiting cases. For $\eta(\tau) \to 0$,
which means very precise data, increasingly accurate numerical implementations are needed to 
search for the spectral function maximizing $Q[\rho]$. 
As it is not possible to achieve infinite numerical precision, the numerical computation will 
eventually become unstable below some value of $\eta$. 
On the other hand, for $\eta(\tau) \to \infty$, the prior probability dominates $Q[\rho]$
and one will simply get $\rho(\omega) = m(\omega)$ as a result of the analysis. It is hence
understood that the error $\eta(\tau)$ should neither be too large nor too small for
the numerical implementation of MEM to work properly. When dealing with very precise or even 
exact data, one therefore has to artificially increase the value for the error $\eta(\tau)$ in order to
avoid numerical problems, see also Sec.~\ref{sec:model_study}.

Let us mention here that the procedure described above is not the only possible
Bayesian method used to reconstruct the spectral function. In particular one can find alternative formulations
of the prior probability, as it was recently proposed in \cite{BurnierRothkopf2013}.
Also, for an alternative application of MEM not to Euclidean time data, but to QCD sum rules, see \cite{GublerOka2010}. 

\subsection{Backus-Gilbert method}
\label{sec:BG}

The Backus-Gilbert (BG) method \cite{BrandtBG} starts by defining an estimator of the spectral function,  
 \begin{equation} \label{BG:ConvolutionDeltaFunction}
     \bar{\rho}(\omega_0) = \int^{\infty}_0 d\omega \, \delta(\omega_0,\omega) \rho(\omega).
 \end{equation}
 The estimator $\bar{\rho}(\omega_0)$ is the convolution of the exact spectral function $\rho(\omega)$ with a normalized resolution function $\delta(\omega_0,\omega)$,
   \begin{equation}\label{BG:norm}
  \int^{\infty}_0 d\omega \, \delta(\omega_0,\omega) = 1.
 \end{equation}  
The BG method is set up as a linear problem which simplifies the interpretation of results and error estimates. To achieve this, the resolution function is first expressed as a linear combination of the kernel profiles taken at different discrete values $\tau_i$ of the Euclidean time,
 \begin{equation}
   \delta(\omega_0,\omega) = \sum_{i} q_i(\omega_0) K(\tau_i, \omega),
 \end{equation}
 with coefficients $q_i(\omega_0)$ as determined below, and with $K(\tau_i, \omega)$ either being the kernel in the spectral representation (\ref{eq:Lehmann:tau})  or that in a Green-Kubo (GK) relation as the one used for the electrical conductivity in Sec.~\ref{sec:conductivity}. In either case it is defined by a general relation of the form  
  \begin{equation}
   G^E(\tau)= \int_0^\infty d \omega\,  K(\tau, \omega) \rho(\omega).
  \end{equation}
Due to the linearity of this relation, one obtains
 \begin{equation} \label{BG:Estimator}
   \bar{\rho}(\omega_0) = \sum_i q_i(\omega_0) G^E(\tau_i).
 \end{equation}
 The coefficients $q_i(\omega_0)$ in Eq.~(\ref{BG:Estimator}) are then
 determined by minimizing the frequency resolution  $D(\omega_0)$ around $\omega_0$, here defined as the second moment of the square of the resolution function,
 \begin{equation}
   D(\omega_0) \equiv \int^{\infty}_0 d\omega \, (\omega-\omega_0)^2 \delta^2(\omega_0,\omega) 
 \end{equation}
while maintaining its normalization condition in Eq.~(\ref{BG:norm}) as a constraint. The result of this minimization yields \cite{BrandtBG}
 \begin{equation}
   q_i(\omega_0) =\frac{W(\omega_0)^{-1}_{ij}R_j}{R_m W(\omega_0)^{-1}_{mn}R_n},
 \end{equation}
 where 
 \begin{equation} \label{BG:W}
   W(\omega_0)_{ij} = \int^{\infty}_0 d\omega \, (\omega-\omega_0)^2 K(\tau_i,\omega) K(\tau_j,\omega),
 \end{equation}
 and
\begin{equation} \label{BG:R}
 R_i = \int^{\infty}_0 d\omega \, K(\tau_i,\omega). 
 \end{equation} 
If the behavior of the kernel at large or small frequencies leads to divergences in the integrals in Eqs.~(\ref{BG:W}) or (\ref{BG:R}), simple reweighting can help: one rescales both, the spectral function and the kernel in Eq.~(\ref{eq:Lehmann:tau}) by reciprocal factors in a way suitable to improve the convergence of the integrals in Eq.~(\ref{BG:W}).   
The matrix $W$ is usually extremely ill-conditioned, with a condition number $C(W) \equiv \lambda_{\text{max}}/{\lambda_{\text{min}}}$, the maximal ratio of eigenvalues of the matrix $W$, of the order of $10^{20}$ or so. Therefore, some regularization is necessary to obtain sensible results for a given set of data $G^E(\tau_i)$.
 
Previous studies employing the BG method have used the so-called ``covariance" regularization \cite{BrandtBG}. In this approach, the following modification is made in Eq.~(\ref{BG:W})
\begin{equation}
  W(\omega_0)_{ij} \to (1-\lambda) W(\omega_0)_{ij} + \lambda C_{ij}, 
\end{equation}
where $\lambda$ here is a small regularization parameter and $C_{ij}$ is the covariance matrix of the Euclidean correlator $G^E$ as before. This replacement improves the condition number of the matrix $W$ while the width of the resolution functions in frequency space can still stay relatively small. Increasing $\lambda $ generally improves the stability of the inversion at the expense of resolution.

Although covariance regularization typically performs quite well, one might wonder about the merits of other commonly used regularization methods for ill-posed problems. Furthermore, in numerical studies where a covariance matrix cannot be constructed (i.e.~when the Euclidean correlator data are obtained using a non-stochastic procedure), covariance regularization cannot be applied.  Here, we employ an alternative method known as the Tikhonov regularization technique \cite{Tikhonov} which is widely used for ill-posed problems of the form $Ax=b$. In this method, one seeks a solution to the modified least-squares function
\begin{equation}
   \text{min} \left( \Vert A x - b \Vert^2_2 + \Vert \Gamma x \Vert^2_2 \right),
   \label{BG:Tikh_def}
\end{equation}
where $\Gamma$ is an appropriately chosen matrix.  The effect of various types of Tikhonov regularizations on the matrix $W$ can be most easily seen by employing the singular value decomposition (SVD). In this procedure 
\begin{equation}
   W = U \Sigma V^{\top}, ~UU^{\top} = VV^{\top} = {\bm 1},
\end{equation}
where $\Sigma = \text{diag}(\sigma_1,\sigma_2,\dots,\sigma_N),~\sigma_1 \geq \sigma_2 \geq \cdots \geq \sigma_N$. The inverse is thus easily expressed as 
\begin{equation}
  W^{-1} = V D U^{\top},~D = \text{diag}( \sigma^{-1}_1,\sigma^{-1}_2,\dots,\sigma^{-1}_N ). 
\end{equation}
In the standard Tikhonov regularization, one modifies the matrix $D$ in the following way 
\begin{equation} \label{BG:SVDStandardTikhonov}
   D_{ij} = \frac{\delta_{ij}}{\sigma_i} \to \tilde{D}_{ij} = \delta_{ij}\, \frac{\sigma_i}{\sigma^2_i + \lambda^2},
\end{equation}
where $\lambda$ is again the regularization parameter. One can see that the singular values which satisfy $\sigma_i \ll \lambda $ are smoothly cut off. This procedure corresponds to $\Gamma = \lambda {\bm 1}$ in Eq.~(\ref{BG:Tikh_def}). One thus pays a price for solutions that are not ``smooth". In general, for small $\lambda$, the solutions fit the data well but are oscillatory, while at large $\lambda$, the solutions are smooth but do not fit the data as well.
The error estimation can be done using data binning (the procedure is described in  \cite{Hubbard_model}) or directly using Eq.~(\ref{BG:Estimator}) together with the covariance matrix $C_{ij}$ for the Euclidean correlator data $G^E(\tau_i)$.

The algorithm for choosing the optimal value of $\lambda$ is based on the ``global relative error" for the spectral function
\begin{equation} \label{BG:GlobalRelativeError}
\mathcal{G} \equiv \frac{1}{N_0} \sum_{\omega_0} \frac{\sigma\left( \bar{\rho}(\omega_0) \right)}{\bar{\rho}_{\text{avg}}(\omega_0)}, 
\end{equation} 
where the sum in the above expression runs over the centers of the resolution functions and $N_0$ is the  number of resolution functions with different centers $\omega_0$. Our basic criteria for the choice of $\lambda$ is that the ``global relative error" should be within the interval  $5-10\%$. We thus sufficiently suppress the statistical error while still maintaining good resolution in frequency space. 

Using the quantity defined in Eq.~(\ref{BG:GlobalRelativeError}) as a measure, we start from small $\lambda$ and increase it until we have obtained the desired statistical error. Typically, we have taken $\lambda=10^{-9}-10^{-8}$ in obtaining the results in this paper. 

\subsection{Schlessinger point method}
\label{sec:Pade}

The Schlessinger point (SP) or Resonances Via Pad\'{e} (RVP) method \cite{Schlessinger1966, TripoltHaritanWambachEtAl2016} is based on a rational-fraction representation similar to Pad\'{e} approximation methods.

Given a set of $N$ input points $(x_i, y_i)$ one first constructs a continued fraction of the form,
\begin{equation}
C_N(x) = \cfrac{y_1}{1+\cfrac{a_1(x-x_1)}{1+\cfrac{a_2(x-x_2)}{\vdots\dots\, a_{N-1}(x-x_{N-1})}}},
\label{eq:pade2}
\end{equation}
where the coefficients $a_i$ are chosen such that the function $C_N(x)$ acts as an interpolation through all the points, i.e.
\begin{equation}
C_N(x_i)=y_i,\,\, i= 1,2,\dots, N.
\end{equation}
The coefficients $a_i$ can easily be determined recursively, see \cite{Schlessinger1966} for details. Errors of the input data can be taken into account by repeating this procedure many times for different sets of input points, $(x_i, y_i\pm \epsilon_i)$, where $\epsilon_i$ can be chosen randomly within the given statistical error. The results obtained in this way can be used to define an error for the final result, see also \cite{TripoltHaritanWambachEtAl2016}.

The function $C_N(x)$ can also be expressed as a rational fraction
\begin{equation}
C_N(x) = P(x)/Q(x),
\end{equation}
where $P$ and $Q$ are polynomials of order $(N-1)/2$ ($P$ and $Q$) for an odd number of input points and of order $N/2-1$ ($P$) and $N/2$ ($Q$) for an even number of input points. Once the function $C_N(x)$ is determined, its analytic continuation is defined as the meromorphic $C_N(z)$ obtained by replacing the originally real $x$ by the  complex variable $z$. We note that the number of input points will always be chosen around $N\approx 50$ in this work. For fewer input points the reconstruction may depend strongly on the number of points while a larger number may give rise to numerical problems due to a loss of accuracy when calculating the coefficients of the continued fraction.

The SP method is applied to $D^E_N(p^0_n)$ in order to obtain its continuous interpolation by a rational fraction which is then analytically continued to approximate, e.g.~the retarded propagator ${D^R(\omega)=-D^E(p^0\rightarrow i\omega-\epsilon)}$ for $\epsilon\rightarrow 0^+$, whose imaginary part yields the spectral function,
\begin{align}
\rho(\omega)=-\frac{1}{\pi}\,\text{Im}D^R(\omega).
\end{align}
In Sec.~\ref{sec:FRG} we will directly use input data on $D^E(p_0)$, so the intermediate step of applying the Fourier transform is not necessary in this case. 

Apart from reconstructing the spectral function, the SP method can also be used to locate resonance poles of propagators in the complex energy plane \cite{TripoltHaritanWambachEtAl2016}.

\section{Model study}
\label{sec:model_study}

\begin{figure}[b]
	\includegraphics[width=0.49\textwidth]{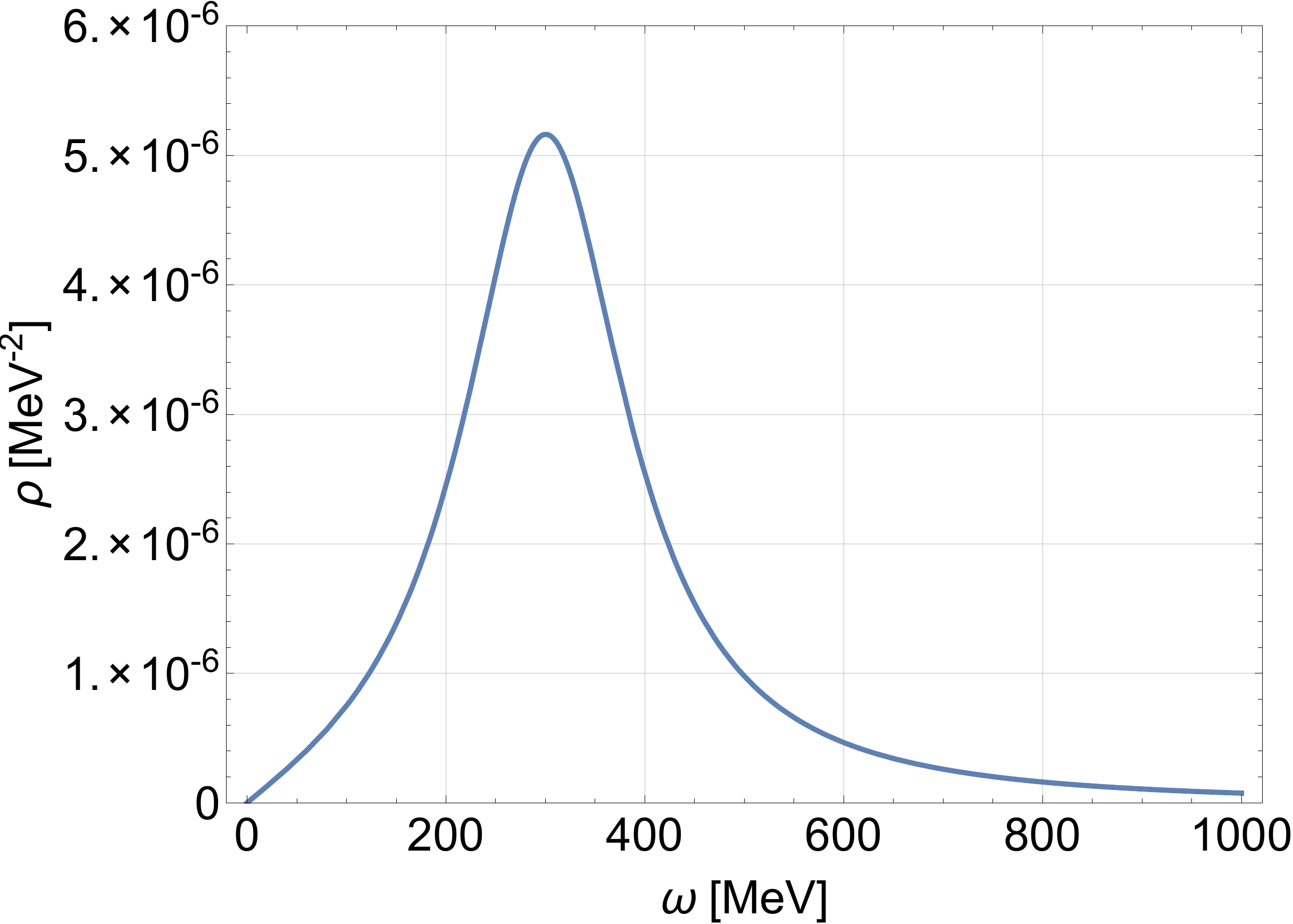}
	\includegraphics[width=0.49\textwidth]{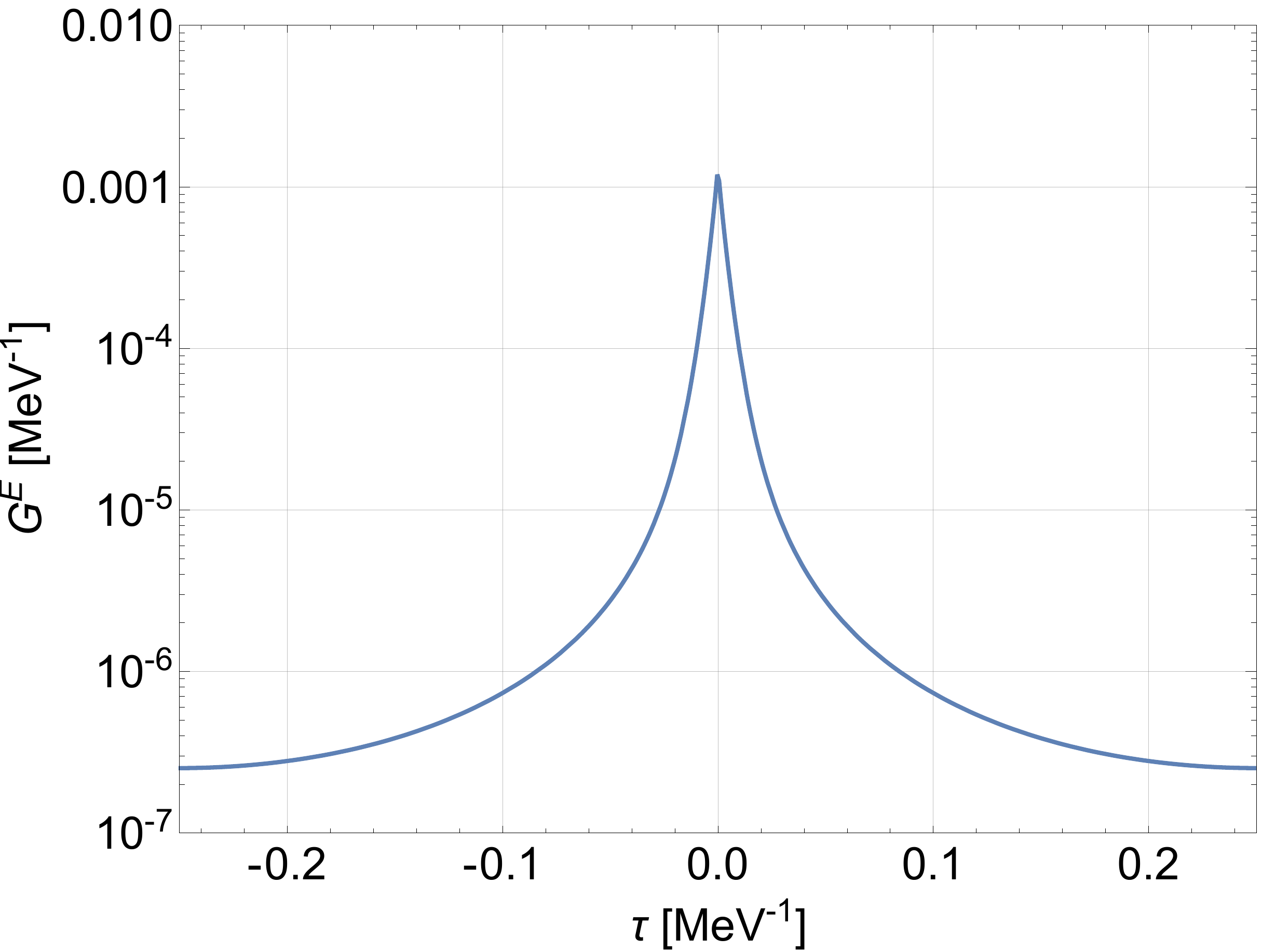}
	\caption{Left: Spectral function given by the Breit-Wigner model, Eq.~(\ref{eq:BW_spectral}). Right: Euclidean correlation function $G^E(\tau)$ as obtained from the Lehmann representation, Eq.~(\ref{eq:Lehmann:tau}).}
	\label{fig:model_spectral}
\end{figure}

In this section we study the following Breit-Wigner type model for a spectral function,
\begin{align}
\label{eq:BW_spectral}
\rho(\omega)&=\frac{1}{\pi}\frac{2\omega\gamma}{(\omega^2-\gamma^2-M^2)^2+4\omega^2\gamma^2},
\end{align}
where we choose $M=300$~MeV, $\gamma=100$~MeV and $T=2$~MeV, see Fig.~\ref{fig:model_spectral}. It fulfills the normalization condition
\begin{align}
1=\int_{0}^{\infty}d\omega \, 2\omega\rho(\omega)
\end{align}
and is positive definite for $\omega>0$. 

In the following we will first study exact input data for $G^E(\tau)$ which we generate by using Eq.~(\ref{eq:Lehmann:tau}) and compare the reconstructed spectral functions obtained from the different methods. Then we will add random noise to $G^E(\tau)$ and vary the number of input points to determine the range of applicability of the different reconstruction methods.

\subsection{Results using exact input data}

\begin{figure}[t!]
	\centering\includegraphics[width=0.49\textwidth]{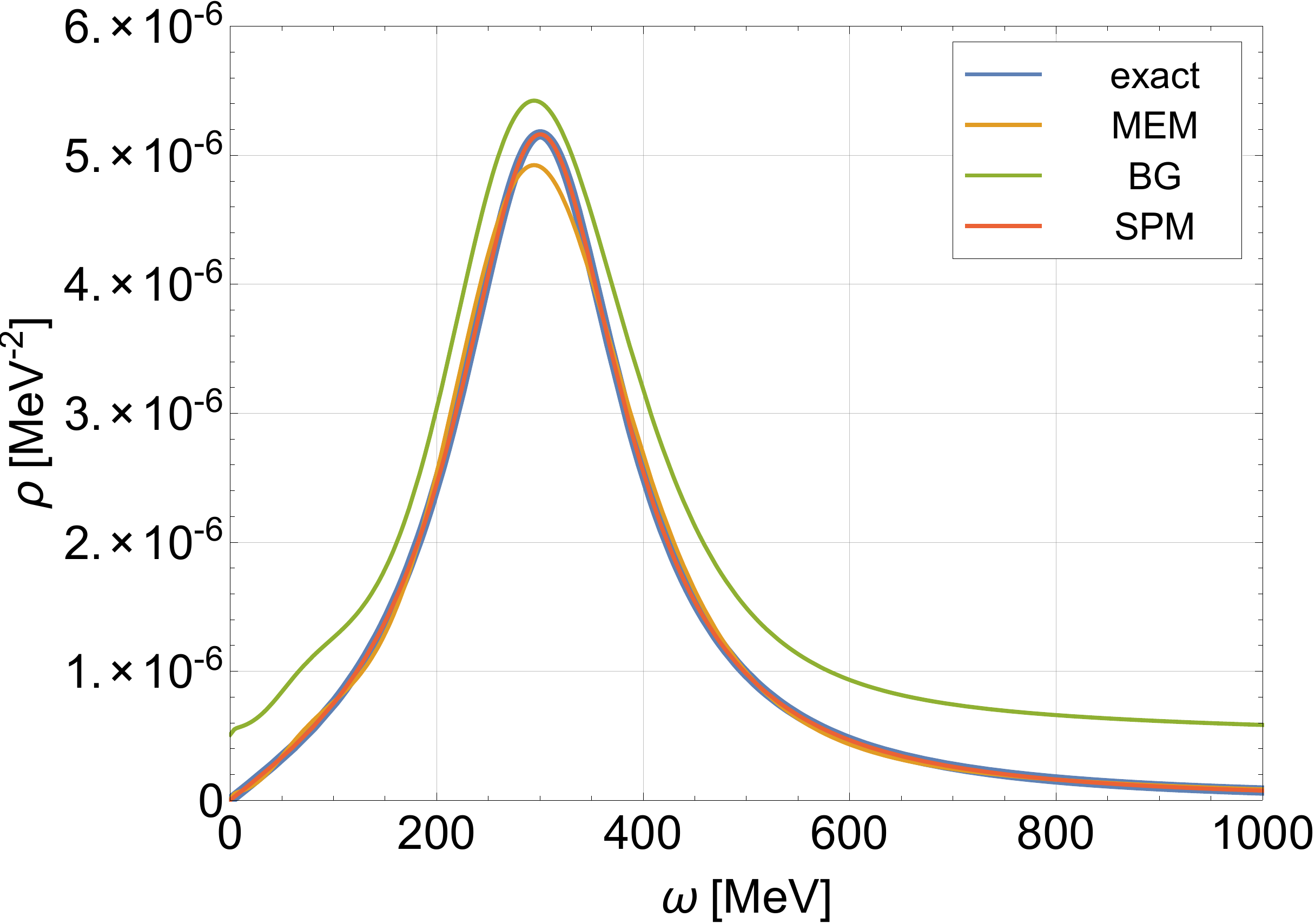}
	\caption{Comparison of the exact model spectral function with the reconstructed spectral function using the Maximum Entropy Method, the Backus-Gilbert method and the Schlessinger point method.}
	\label{fig:model_spectral_exact_comparison}
\end{figure}

As a first benchmark test of the three methods we provide them here with up to $N=2048$ data points on the Euclidean correlation function $G^E(\tau)$ as obtained by Eq.~(\ref{eq:Lehmann:tau}) with negligible error. 
As our numerical implementation of MEM becomes unstable for very small errors, we use here an artificial 
internal error of $\eta(\tau) = 10^{-5} \times G^E(\tau)$. 
As already mentioned earlier, we employ Bryan's algorithm in our code. Furthermore, we 
use double numerical precision and follow the prescription of \cite{AsakawaNakaharaHatsuda2000} for the integration 
routines and other intermediate steps of the calculation. 
A more precise numerical implementation would allow MEM to go to smaller errors.
The results are shown in Fig.~\ref{fig:model_spectral_exact_comparison} together with the exact spectral function. While all methods are in principle able to reconstruct the model spectral function very well, there are some small differences. The BG method seems to overestimate the spectral function for all energies, while the peak height from MEM is a bit too low. The SP method gives the best reconstruction for this simple example and reproduces the spectral function almost exactly.

\subsection{Dependence on number of input points and error}

We will now study how many input points for the Euclidean correlation function are needed and how large their statistical error can be in order for the different methods to produce a good reconstruction of the spectral function. Therefore we add Gaussian noise of the form
\begin{align*}
f(x)=\frac{1}{\sqrt{2\pi}\sigma}e^{-\frac{(x-\mu)^2}{2\sigma^2}}
\end{align*}
to the data, with $\mu=y_i$ and different values for the relative error $\sigma$ which is connected to the absolute error $\eta$ used in MEM by $\sigma(\tau)=\eta(\tau)/G^E(\tau)$. 

\begin{figure}[b!]
	\includegraphics[width=0.49\textwidth]{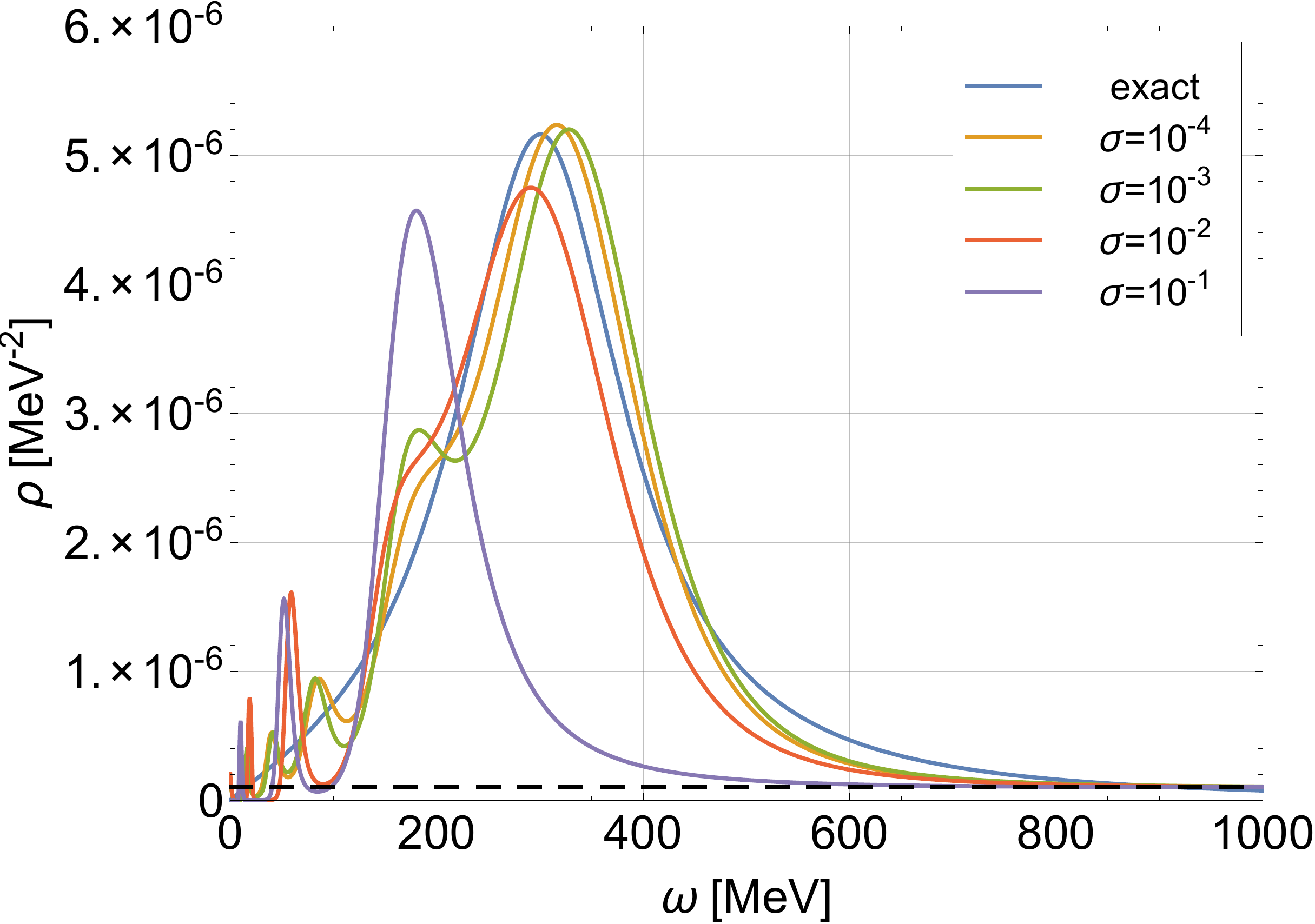}
	\includegraphics[width=0.49\textwidth]{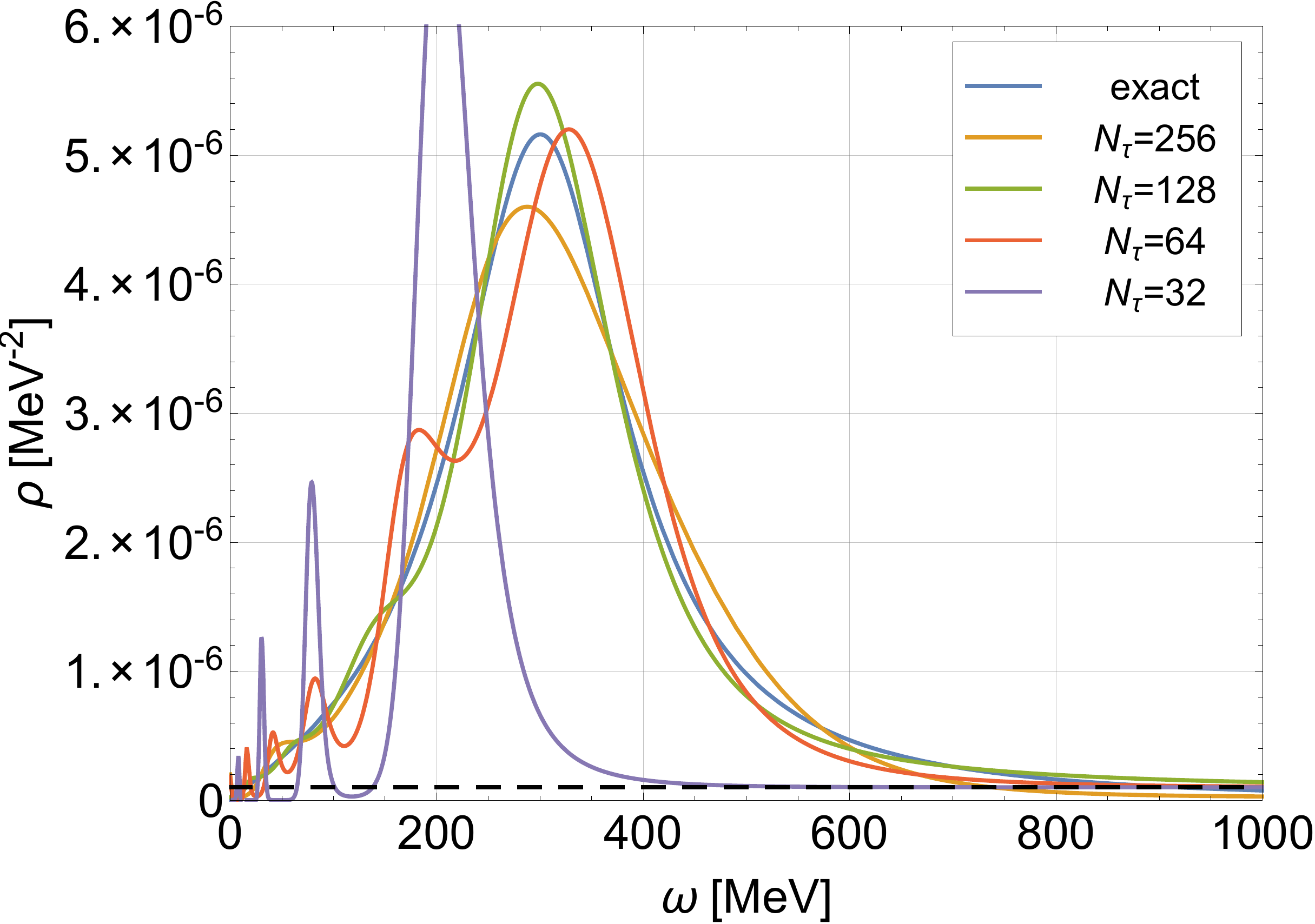}
	\caption{Left: Comparison of the exact model spectral function with the reconstruction obtained from the Maximum Entropy Method for a fixed number of $N_\tau=64$ input points for the Euclidean correlation function $G^E(\tau)$ but different error. The employed default model is denoted as a dashed line. Right: Same as left but for a fixed error of $\sigma=10^{-3}$ and different numbers of input points.}
	\label{fig:spectral_BW_MEM_N64_errors}
\end{figure}

In Fig.~\ref{fig:spectral_BW_MEM_N64_errors} we compare the reconstructed spectral functions obtained from MEM for different numbers of input points and different errors. For a fixed number of $N_\tau=64$ input points the peak position as well as its width is reproduced correctly for $\sigma\leq 10^{-2}$ while the peak is at too small energies for $\sigma= 10^{-1}$. When using a fixed error of $\sigma=10^{-3}$, the peak position and width are recovered very well for $N_\tau\geq 64$ while the peak is shifted to smaller energies for $N_\tau=32$. We also note that MEM always produces a ``ringing", i.e.~artificial oscillations in the spectral function, at small energies. This is a known MEM artifact and makes it difficult to distinguish physical peaks from these unphysical ones in a setting where the exact answer is not known. 
The ringing of the extracted spectra at small $\omega$ is partly caused by the behavior 
of the original spectral function approaching zero in the $\omega \to 0$ limit. This can 
cause trouble for MEM, which assumes $\rho(\omega) > 0$. It is possible to avoid this 
issue in the present case by defining the kernel of Eq.(\ref{eq:Lehmann:tau}) as 
\begin{align}
K(\tau,\omega) = \omega \times \frac{\cosh(\omega(\tau -\beta/2))}{\sinh(\beta\omega/2)},
\end{align}
and thus extracting $\rho(\omega)/\omega$ instead of $\rho(\omega)$. Carrying out such 
an analysis, we have found that for errors larger than $\sigma = 10^{-4}$ it is indeed possible 
to reduce the ringing at small $\omega$, however at the price of distorting the main peak 
away from the exact solution. For $\sigma = 10^{-4}$, the ringing is not reduced for small 
$\omega$, but even enhanced for $\omega > 50$ MeV. 
We will hence in all the MEM analyses shown in this paper use the standard kernel 
\begin{align}
K(\tau,\omega) = \frac{\cosh(\omega(\tau -\beta/2))}{\sinh(\beta\omega/2)},
\end{align}
which performs better for the purpose of extracting the properties of the main peak.

\begin{figure}
	\includegraphics[width=0.49\textwidth]{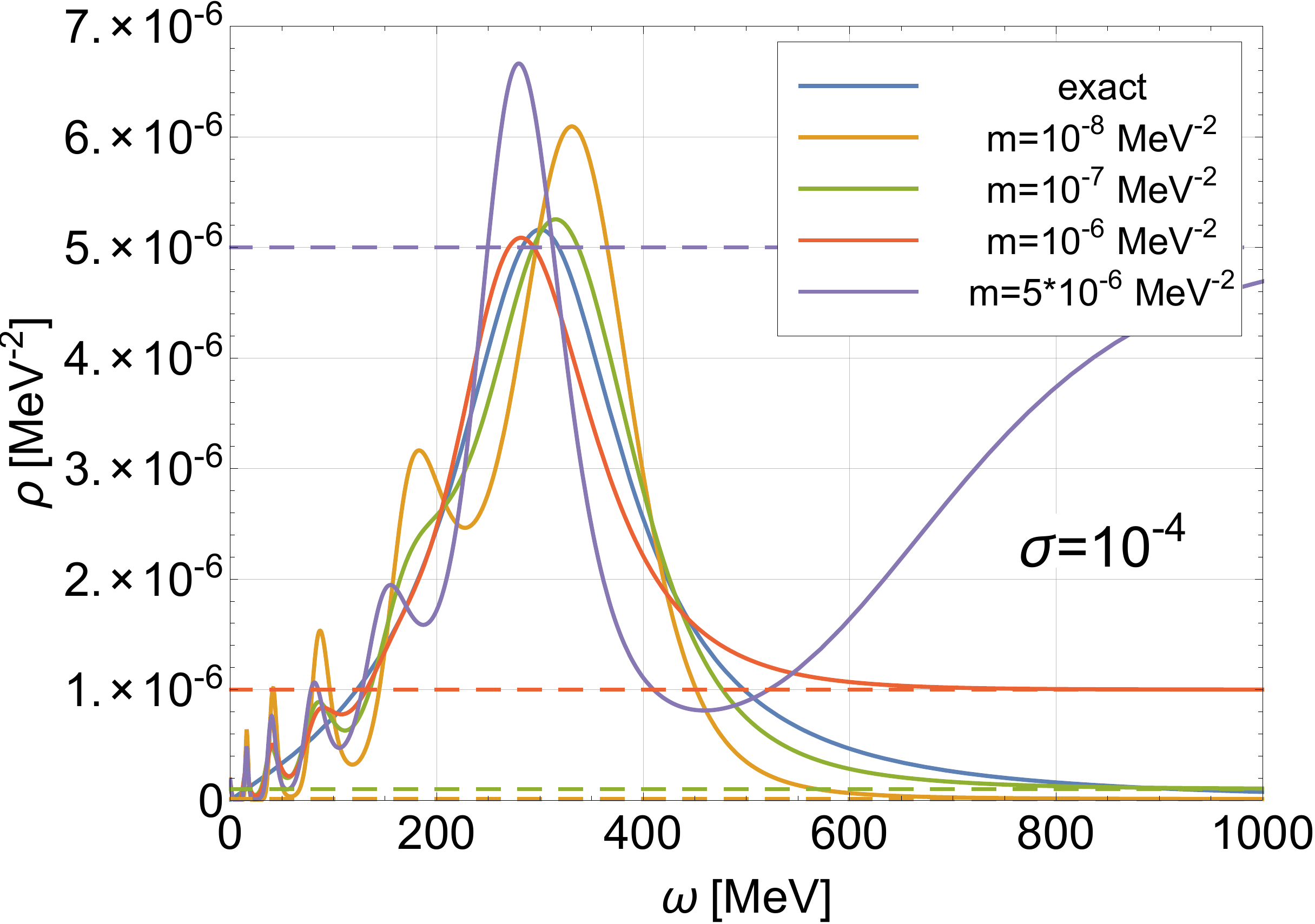}
	\includegraphics[width=0.49\textwidth]{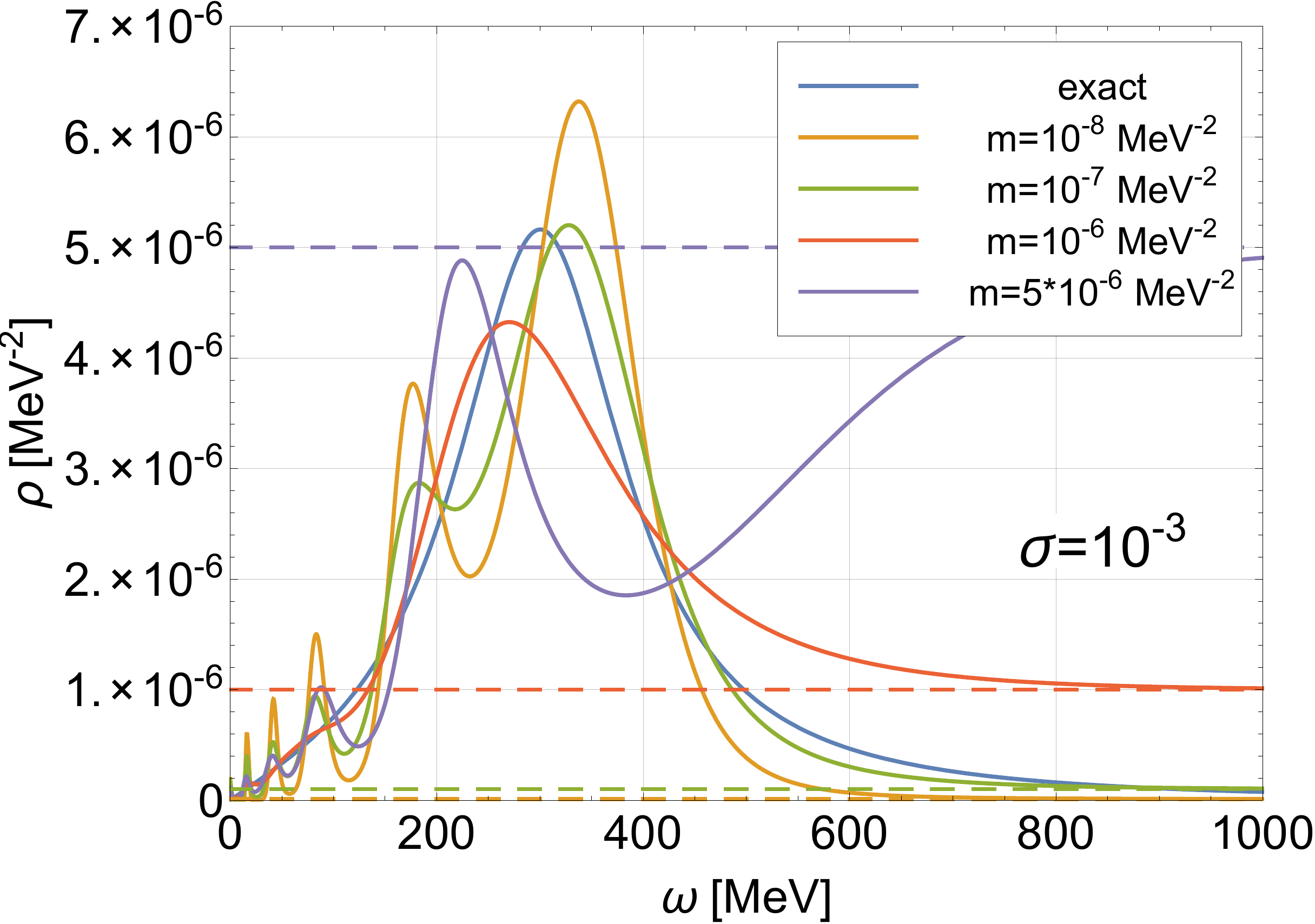}
	\caption{Left: Comparison of the exact model spectral function with the reconstruction obtained from the Maximum Entropy Method for a fixed number of $N_\tau=64$ input points for the Euclidean correlation function $G^E(\tau)$ and fixed error of $\sigma=10^{-4}$ but different default models. The employed default models are denoted as dashed lines. Right: Same as left but for an error of $\sigma=10^{-3}$.}
	\label{fig:spectral_BW_MEM_N64_def_model}
\end{figure}

The extracted spectrum of any MEM analysis generally depends on the choice of the default model 
(see, for instance, \cite{AartsAlltonFoleyEtAl2007,EngelsVogt2010,DingKaczmarekMukherjeeEtAl2018}). 
Let us illustrate this dependence here with a few examples. 
To keep the discussion simple, we here consider only constant default models with varying absolute 
values. On the left plot of Fig.\,\ref{fig:spectral_BW_MEM_N64_def_model}, we show results of analyses carried out under the same conditions 
as the orange curve on the left plot of Fig.\,\ref{fig:spectral_BW_MEM_N64_errors} ($N_{\tau} = 64$ and $\sigma = 10^{-4}$), but with 
changing default models. It is observed that the qualitative features of the main peak around 
$\omega = 300$ MeV remain roughly the same, while the ringing at low $\omega$ depends strongly 
on the choice of the default model. Clearly, both for very large and small values of the default model, we observe 
strong ringing. For $\omega \geq 600$ MeV, the spectral function simply approaches the default 
model, which shows that MEM is not able to make reliable statements about the spectral behavior 
in that region. To examine how the dependence on the default model is affected by the error, we 
show on the right plot of Fig. 4 the results of the same analysis, but with the error increased to 
$\sigma = 10^{-3}$. It is seen in this plot, that while the gross features remain the same, the 
form of the reconstructed main peak starts to depend strongly on the default model, such that 
for some cases it even gets difficult to distinguish between the physical peak and non-physical 
ringing.

The above discussion shows that it is important to choose a suitable default model, which leads to 
minimal ringing and a satisfactory extraction of the spectral structures of interest. In some cases, such as 
QCD, the asymptotic behavior of the spectral function at large $\omega$ is in fact known and provides 
a natural choice for the default model. We will make such a choice in the analysis of lattice QCD 
data in Section 5.

\begin{figure}[t!]
	\includegraphics[width=0.49\textwidth]{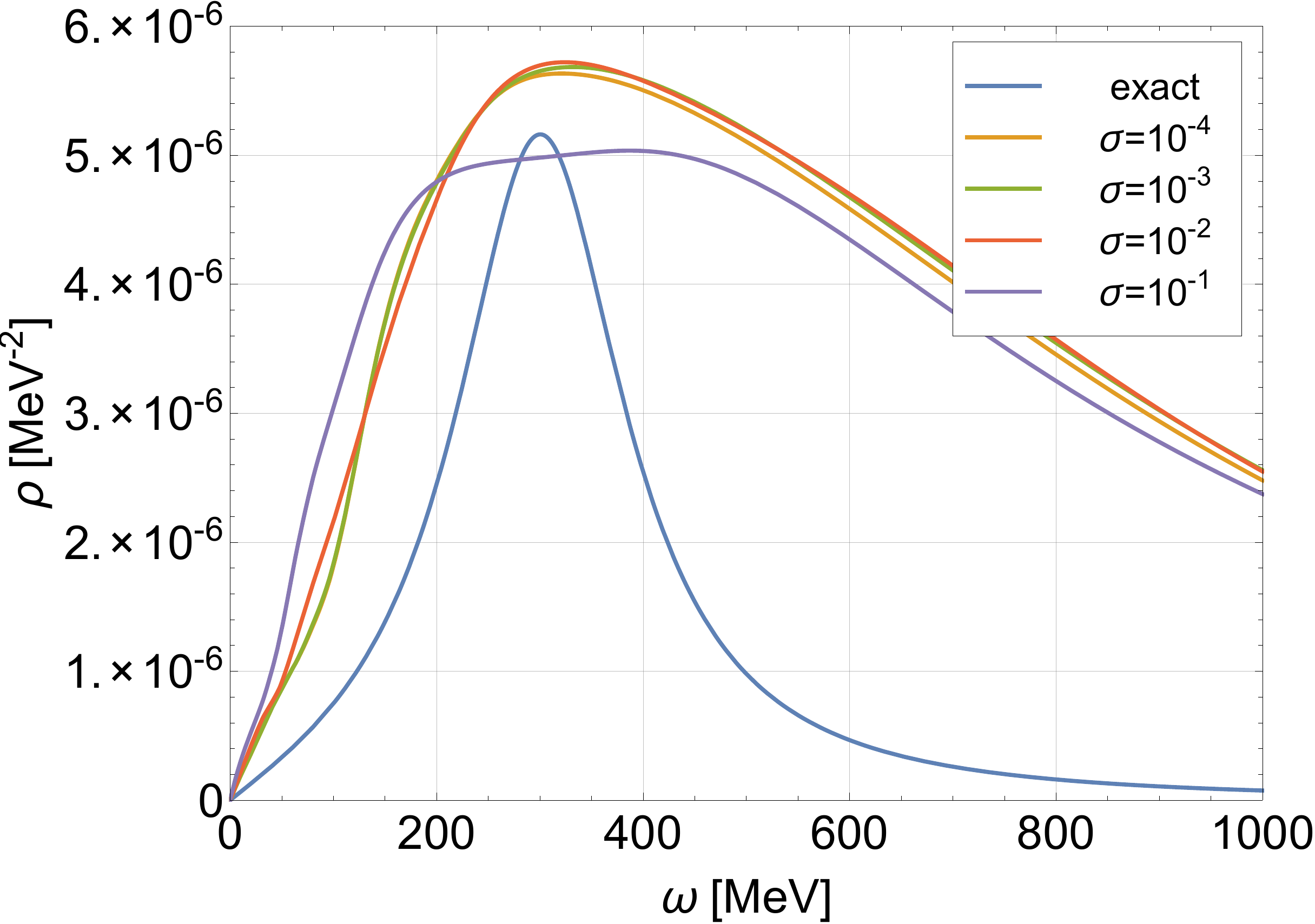}
	\includegraphics[width=0.49\textwidth]{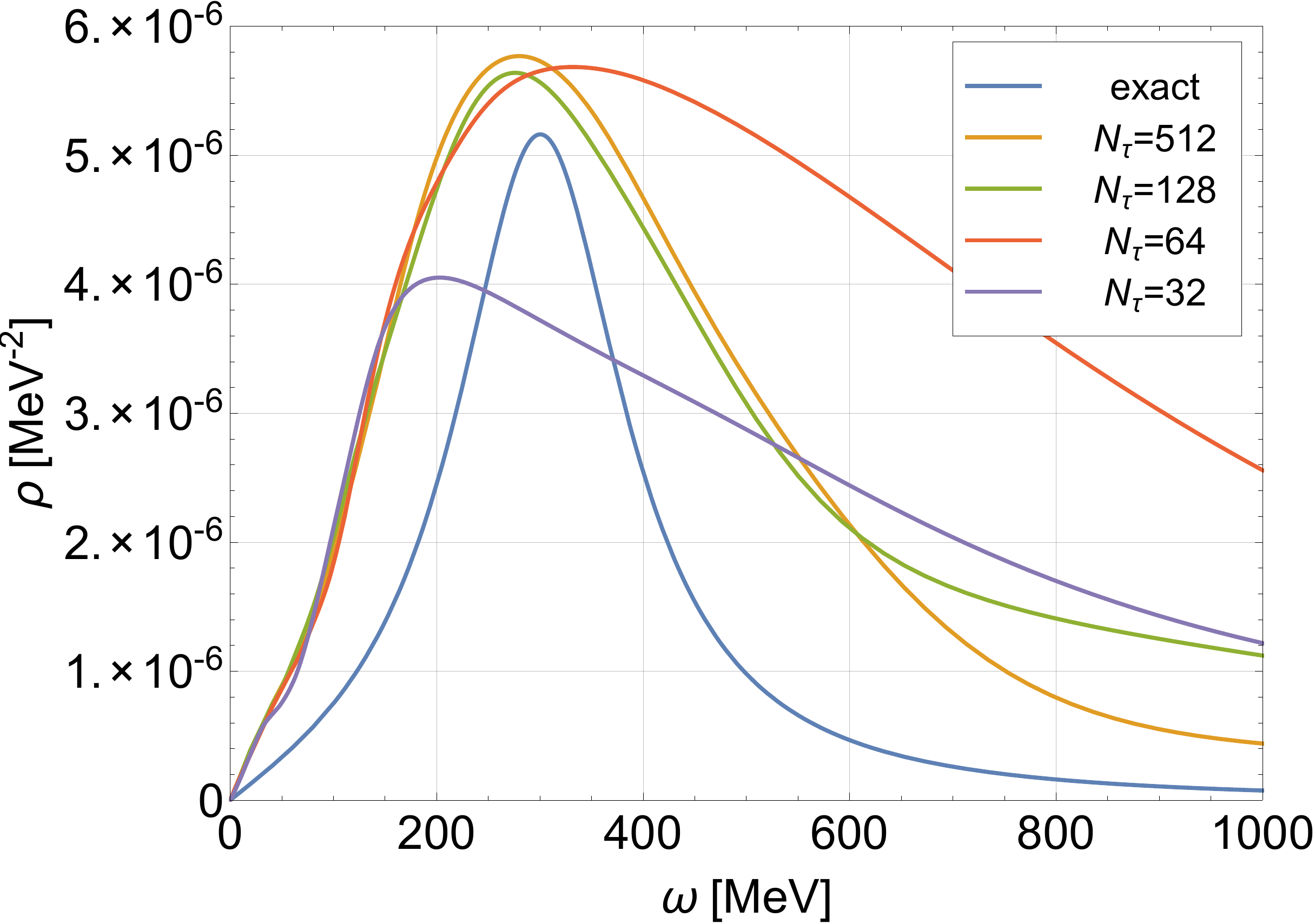}
	\caption{Left: Comparison of the exact model spectral function with the reconstruction obtained from the Backus-Gilbert method for a fixed number of $N_\tau=64$ input points for the Euclidean correlation function $G^E(\tau)$ but different error. Right: Same as left but for a fixed error of $\sigma=10^{-3}$ and different numbers of input points.}
	\label{fig:spectral_BW_BG_N64_errors}
\end{figure}

In Fig.~\ref{fig:spectral_BW_BG_N64_errors} we compare the reconstructed spectral functions obtained from the BG method. For a fixed number of $N_\tau=64$ input points only the peak position is reproduced well for $\sigma\leq 10^{-2}$ while the width is too large and also the shape of the peak is distorted. When using a fixed error of $\sigma=10^{-3}$, the peak position is recovered for $N_\tau\geq 64$ and also the width and the shape of the peak can be obtained better when increasing the number of input points. The effects of the finite resolution are clearly visible, however, which makes the overall quality of the reconstruction appear lower than for MEM.

\begin{figure}[b!]
	\includegraphics[width=0.49\textwidth]{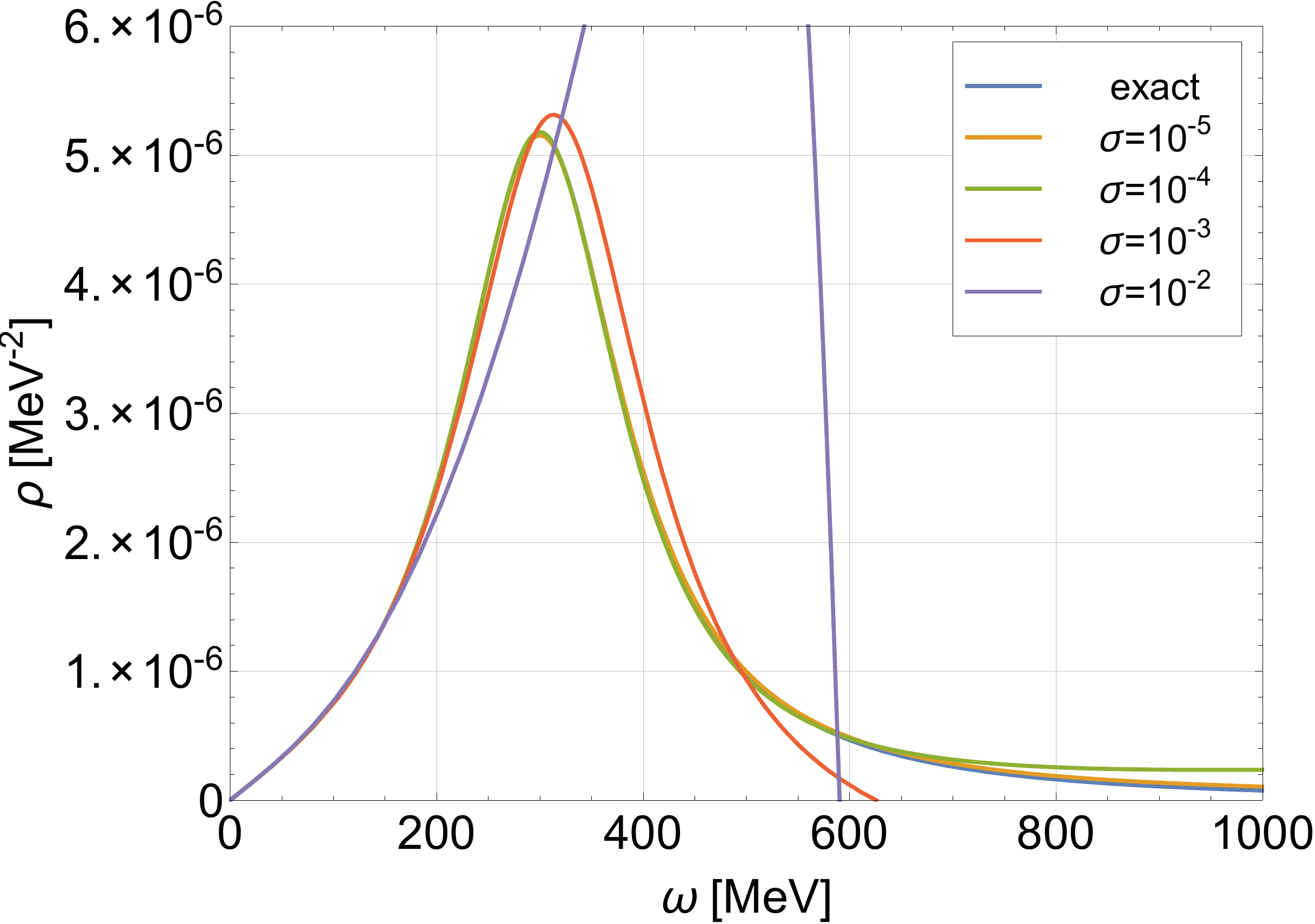}
	\includegraphics[width=0.49\textwidth]{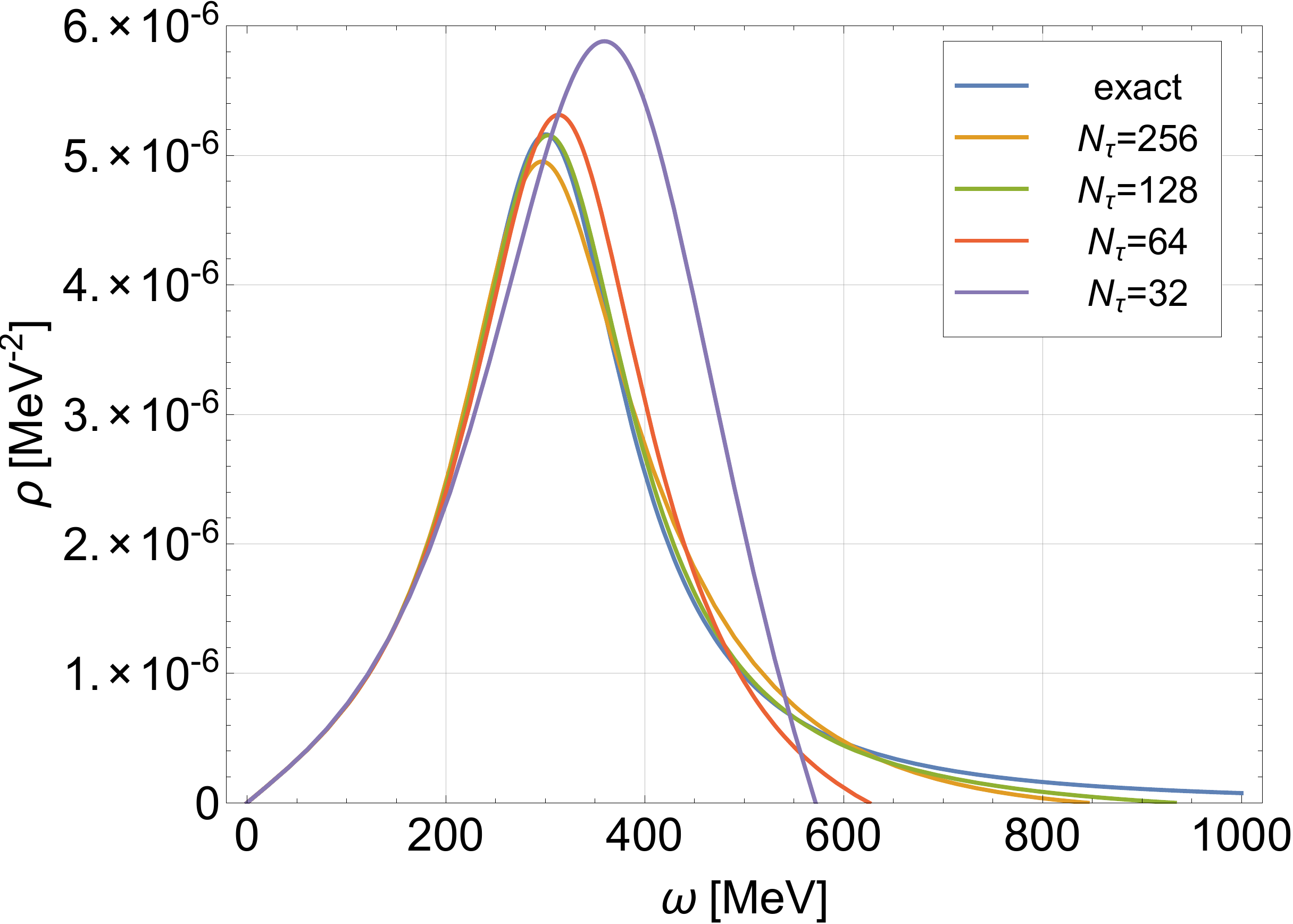}
	\caption{Left: Comparison of the exact model spectral function with the reconstruction obtained from the Schlessinger point method for a fixed number of $N_\tau=64$ input points for the Euclidean correlation function $G^E(\tau)$ but different errors. Right: Same as left but for a fixed error of $\sigma=10^{-3}$ and different numbers of input point.}
	\label{fig:spectral_BW_RVP_N64_errors}
\end{figure}

In Fig.~\ref{fig:spectral_BW_RVP_N64_errors} we show the spectral functions obtained from the SP method. For a fixed number of $N_\tau=64$ input points the spectral function is reconstructed almost exactly for $\sigma\leq 10^{-3}$ while for $\sigma\geq 10^{-2}$ the reconstruction breaks down. We note that it is also possible that the reconstructed spectral functions become negative, as seen for $\sigma\geq 10^{-3}$ for large energies. When using a fixed error of $\sigma=10^{-3}$, the spectral function is very well reconstructed for $N_\tau\geq 64$ while for $N_\tau=32$ the quality of the reconstruction starts to decrease.

\begin{figure}[t]
	\includegraphics[width=0.49\textwidth]{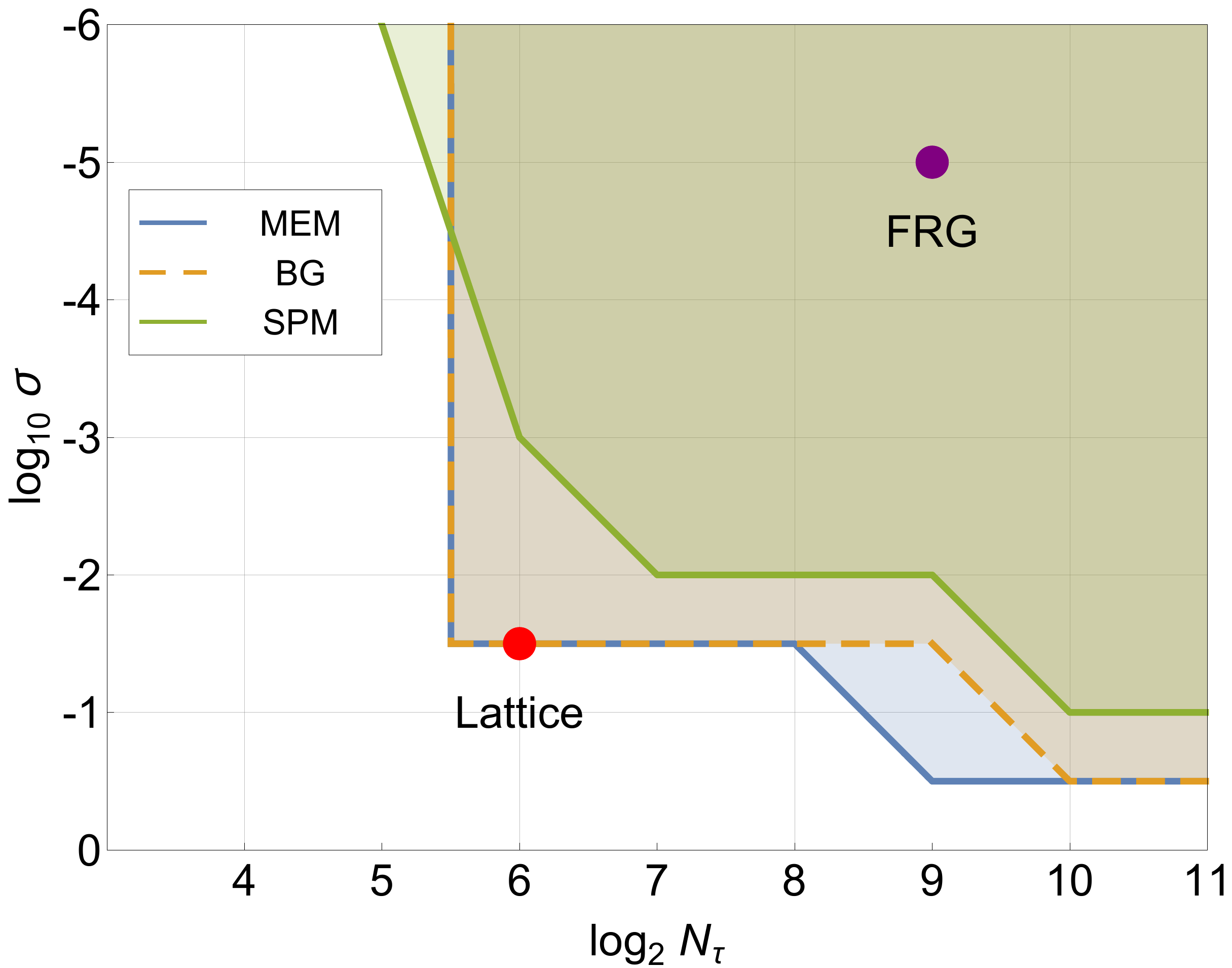}
	\includegraphics[width=0.49\textwidth]{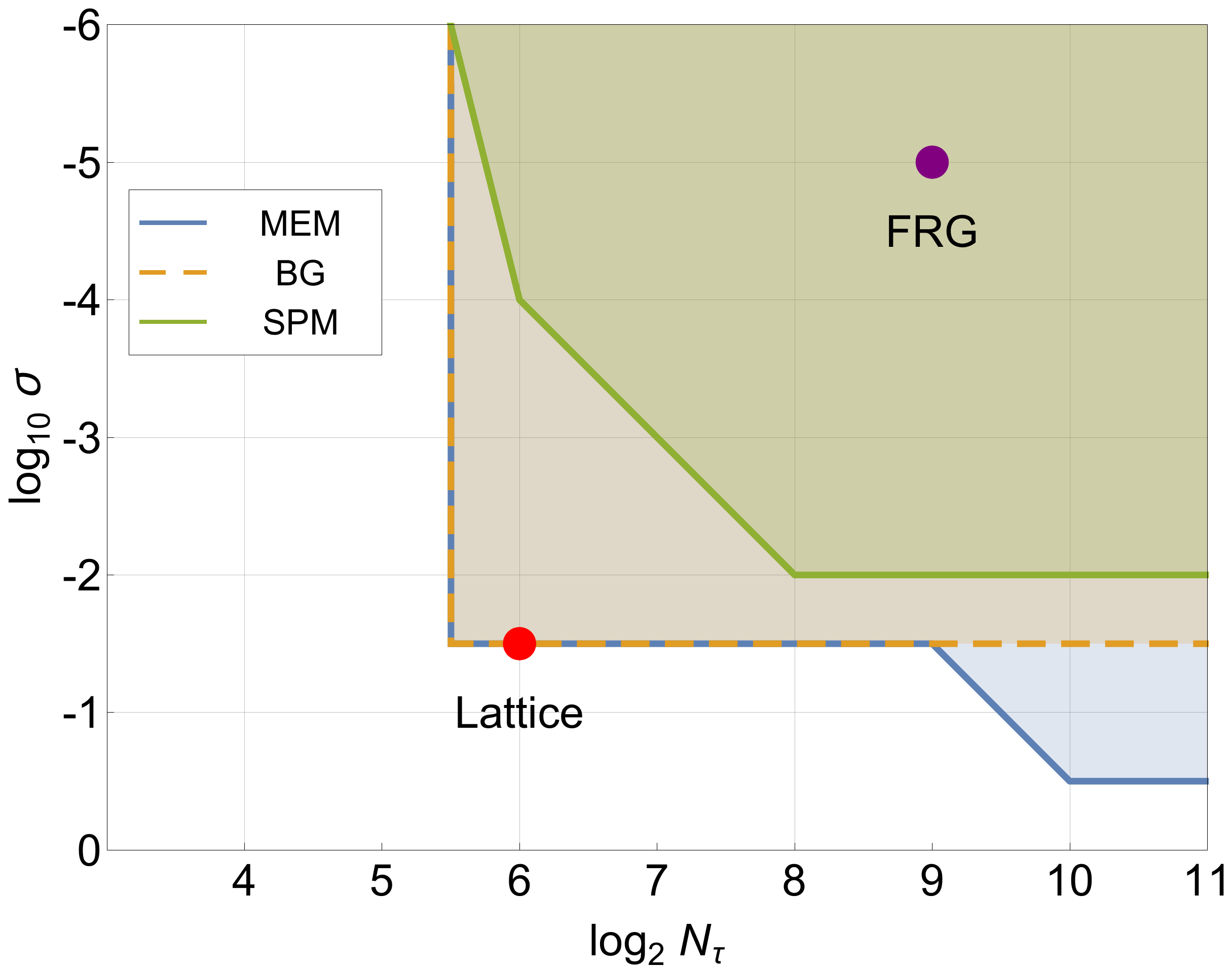}
	\caption{Regime of applicability for the Maximum Entropy Method, the Backus-Gilbert method and the Schlessinger point method in dependence on the number of input points for $G^E(\tau)$ and the relative error of the data with parameter $\sigma$ for a $\tau$-independent error (left) and an error that increases with $\tau$ (right), see text for details. The red dot represents the lattice data analyzed in this work and the purple dot the FRG data.}
	\label{fig:regime_applicability}
\end{figure}

By repeating this analysis for all methods with different numbers of input points and different errors we are able to construct a regime of applicability which is shown on the left-hand side of Fig.~\ref{fig:regime_applicability}. This plot shows for which combinations of input points $N_\tau$ and errors $\sigma$ the individual methods work well, i.e.~when the peak position is reconstructed correctly within a $10\%$ error. We note that this plot is only valid for the particular model studied here. However, we believe that some of the qualitative aspects also hold true for more complicated situations. For example, the regime of applicability of MEM and the BG method are very similar and larger than the one for the SP method. In particular for larger errors, $\sigma\geq 10^{-2}$, the SP method breaks down. However, in the regime where the SP method works, it can be expected to give better reconstructions than MEM or the BG method. This is for example the case for the FRG data studied in the next section, which is marked by a purple dot in Fig.~\ref{fig:regime_applicability}. On the other hand, for the lattice data studied in this work the errors are too large for the SP method, but  also MEM and the BG method reach their limits of applicability by the $10\%$ criterion applied here.

We also note that all methods fail to reconstruct the model spectral function when the number of input points becomes smaller than $N\approx 2^5=32$, independent of the error. The existence of such a natural limit can be easily understood as follows: If the number of input points decreases, the Euclidean time step between the data points, $\Delta \tau = \beta/N$, increases. Since the correlation function rapidly decreases with $\tau$, it will be of the same magnitude as the error already after the first few time steps, unless one has data with exponentially small errors. This makes the reconstruction of the peak in the spectral function extremely difficult.

As already mentioned in Sec.~\ref{sec:MEM}, on the other hand, MEM runs into numerical difficulties when the error of the input data points becomes too small. When exactly this happens depends on the particular implementation of the MEM code and the input data. In the present example this critical value is around $\sigma=10^{-5}$. When analyzing more precise data we therefore have to artificially increase the absolute error $\eta$ in the MEM code. In other words, the reconstructed spectral function obtained from MEM will not become more precise when decreasing the error $\sigma$ any further because we then have to increase the error parameter $\eta$ used in MEM in order to avoid numerical problems.

So far we have discussed a statistical error $\sigma$ that does not depend on the Euclidean time $\tau$. While this is a realistic situation for numerical calculations as discussed in Sec.~\ref{sec:FRG}, it is not the case for the lattice QCD data in Sec.~\ref{sec:Lattice} where the error increases with $\tau$. On the right-hand side of Fig.~\ref{fig:regime_applicability} we therefore show the regime of applicability of the different reconstruction methods where we used an error $\sigma(\tau)$ that increases linearly from $\sigma(\tau=0)=\sigma_0$ to $\sigma(\tau=\beta/2)=100\sigma_0$ where $\sigma_0$ is simply denoted as $\sigma$ in the plot. We observe that the regimes of applicability become slightly smaller, as expected. The effect of the increasing error $\sigma(\tau)$ is, however, not too strong since the data points at small $\tau$, which are the most important ones in the numerical reconstruction, are only mildly affected by this linearly increasing error.

\section{FRG data}
\label{sec:FRG}

In this section we will apply the three analytic continuation methods to numerical Euclidean data obtained from a Functional Renormalization Group (FRG) calculation. The FRG is a non-perturbative continuum framework formulated in Euclidean space-time and is used in particular in statistical physics and QFT, see \cite{Berges:2000ew,Polonyi:2001se,Pawlowski:2005xe,Schaefer:2006sr,Kopietz2010,Braun:2011pp,Gies2012} for reviews. In \cite{Tripolt2014, Tripolt2014a, TripoltSmekalWambach2017} the quark-meson model has been studied within this framework in order to calculate mesonic spectral functions at finite temperature. The new development there was that the so-called FRG flow equations for the corresponding correlation functions, which are originally formulated in Euclidean spacetime, have been analytically continued to real frequencies before they were solved. These analytically continued (aFRG) flow equations therefore provide a non-perturbative functional framework for the direct computation of spectral functions, without the need of any numerical reconstruction method. In the following we will use this theoretical setup to calculate not only these aFRG spectral functions but also the corresponding Euclidean propagators $D^E(p_0)$ of the sigma meson and the pion at the discrete Matsubara frequencies corresponding to the different temperatures. We will then use the numerical data on the Euclidean propagators as input for the three analytic continuation methods and compare the reconstructed spectral functions with the ones from the aFRG flow equations.

In order to check that the aFRG spectral functions $\rho(\omega)$ are in fact consistent with the Euclidean propagators $D^E(p_0)$, we use the Lehmann representation,
\begin{align}
\label{eq:Lehmann_p0}
D^E(p_0)&=\int_{-\infty}^{\infty}d\omega\frac{\rho(\omega)}{\omega+ip_0},
\end{align}
and first compare the result with the numerically obtained Euclidean propagator, see for example Fig.~\ref{fig:Pion_T2_Lehmann}. Since the aFRG spectral functions are only known up to some maximum frequency $\Lambda\approx 1.5$~GeV which is related to the employed UV cutoff, an extrapolation is used to account for the contributions from higher frequencies in Eq.~(\ref{eq:Lehmann_p0}). Although the integral converges, this can result in small numerical differences between the propagator obtained from the Lehmann representation, defined for continuous Euclidean $p_0$, and that from the Euclidean FRG calculation at the discrete Matsubara frequencies. In general, however, the extrapolation can be chosen such that this difference becomes negligible. 

In the following we will always use $N=2048$ points for the Euclidean propagator $D^E(p_0)$ at Matsubara frequencies $p_0=2\pi n T$ with $n\in [-1024,1023]$. Since MEM and the BG method use data on the Euclidean correlation function $G^E(\tau)$ as input, we will perform a discrete Fourier transform of these data, cf.~Sec.~\ref{sec:methods}. The large number of points for $D^E(p_0)$ then ensures that the error introduced by the discrete Fourier transform is small and does not affect the reconstruction. For the Schlessinger point method on the other hand we can directly use the data points on $D^E(p_0)$ from which we select a subset of $N\approx50$ points to reconstruct the spectral function. We also note that for MEM we always use an error parameter of $\eta(\tau)=0.001\, G^E(\tau)$ 
and take 25 bases in the Bryan's algorithm search space. We have checked for the finite temperature cases shown in Figs.\,\ref{fig:Pion_T50_reconstruction} and \ref{fig:Sigma_T140_reconstruction}, that doubling the number of bases to 50 does not change the result.

\subsection{Pion at $T=2$~MeV}

\begin{figure}[b!]
	\includegraphics[width=0.49\textwidth]{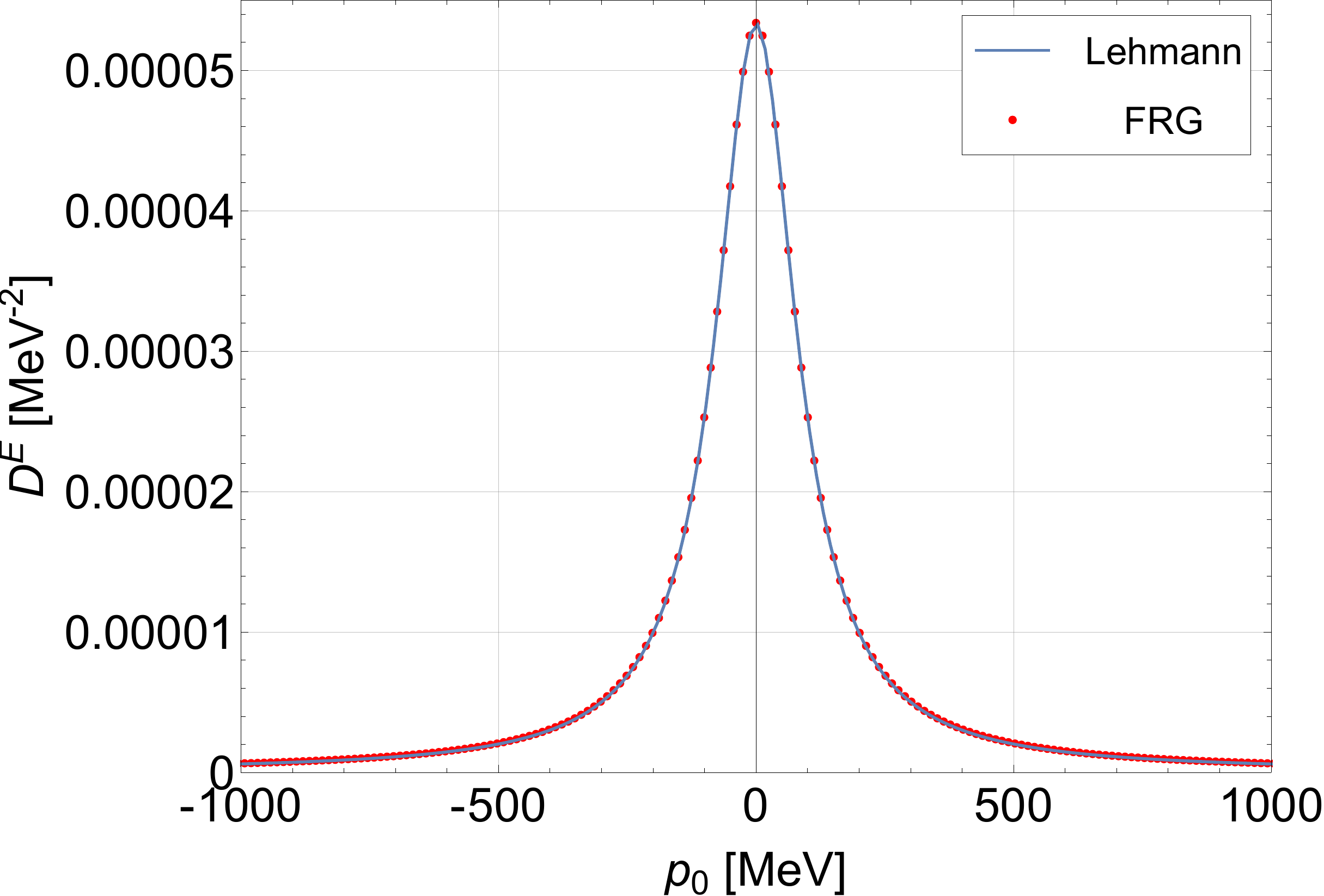}
	\includegraphics[width=0.49\textwidth]{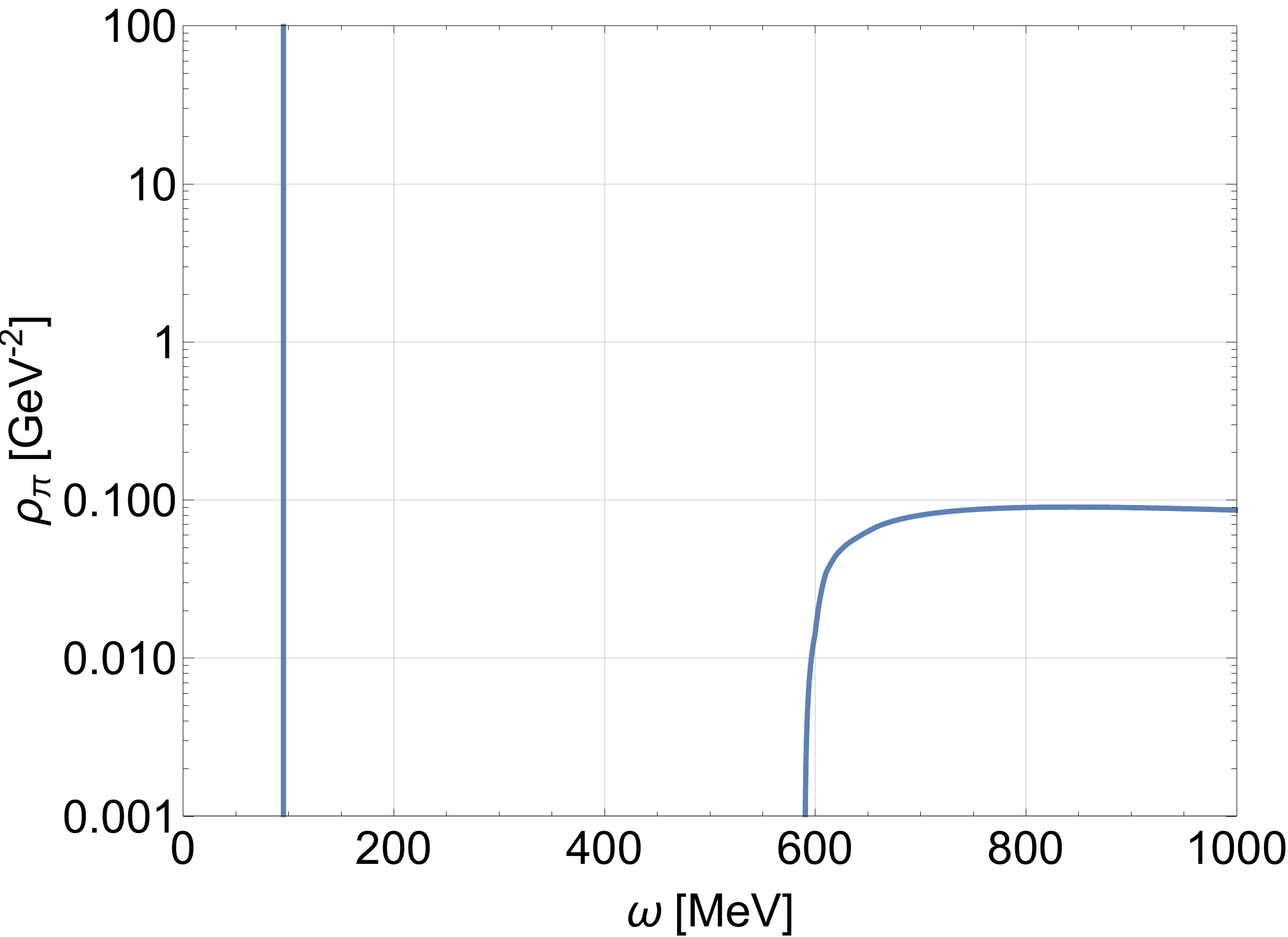}
	\caption{Left: The Euclidean FRG pion propagator $D^E(p_0)$ at the discrete Matsubara frequencies $p_0=2\pi n T$ for $T=2$~MeV compared to that from the Lehmann representation in Eq.~(\ref{eq:Lehmann_p0}). Right: The corresponding input spectral function from the aFRG calculation.}
	\label{fig:Pion_T2_Lehmann}
\end{figure}

\begin{figure}[b!]
	\centering\includegraphics[width=0.49\textwidth]{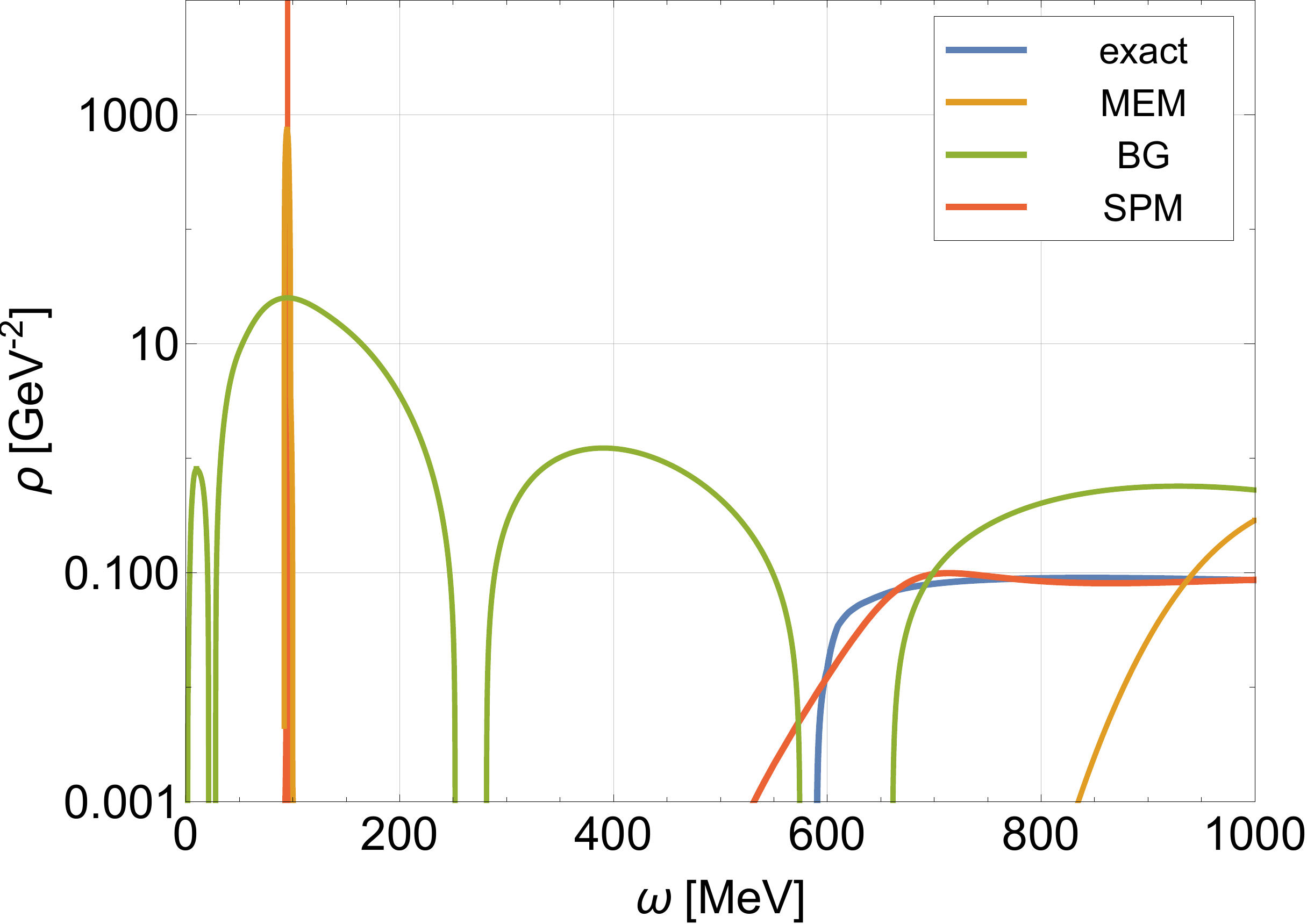}
	\caption{Comparison of the aFRG spectral function for the pion at $T=2$~MeV with the reconstructed spectral functions obtained by using Euclidean data as input for the Maximum Entropy Method, the Backus-Gilbert method and the Schlessinger point method.}
	\label{fig:Pion_T2_reconstruction}
\end{figure}

We will first study the pion propagator and the corresponding spectral function at a low temperature of $T=2$~MeV, see Fig.~\ref{fig:Pion_T2_Lehmann}. It contains a delta peak at $\omega\approx 95$~MeV and a continuum starting at $\omega\approx 600$~MeV. As shown, the data points on $D^E(p_0)$ agree very well with the propagator obtained from the Lehmann representation, which means that the data points obtained from the Euclidean FRG flow equations are consistent with the aFRG spectral function. At low temperatures the spacing between successive Matsubara frequencies, $p_0=2\pi n T$, is smaller than at high temperatures. One therefore has more data points in the interesting frequency range, here of $\omega\in[0,1000]$~MeV, which makes it easier to reconstruct the spectral function for all three analytic continuation methods used here. The reconstructed spectral functions obtained from MEM, the BG method and the SP method are compared in Fig.~\ref{fig:Pion_T2_reconstruction}. The delta peak is well reproduced by the SP method and MEM while the BG method produces a broader peak albeit at the correct location. The continuum part is best reconstructed by the SP method while both MEM and the BG method have more difficulties to capture the onset and the magnitude of the continuum.

\subsection{Sigma at $T=2$~MeV}

\begin{figure}[b!]
	\includegraphics[width=0.49\textwidth]{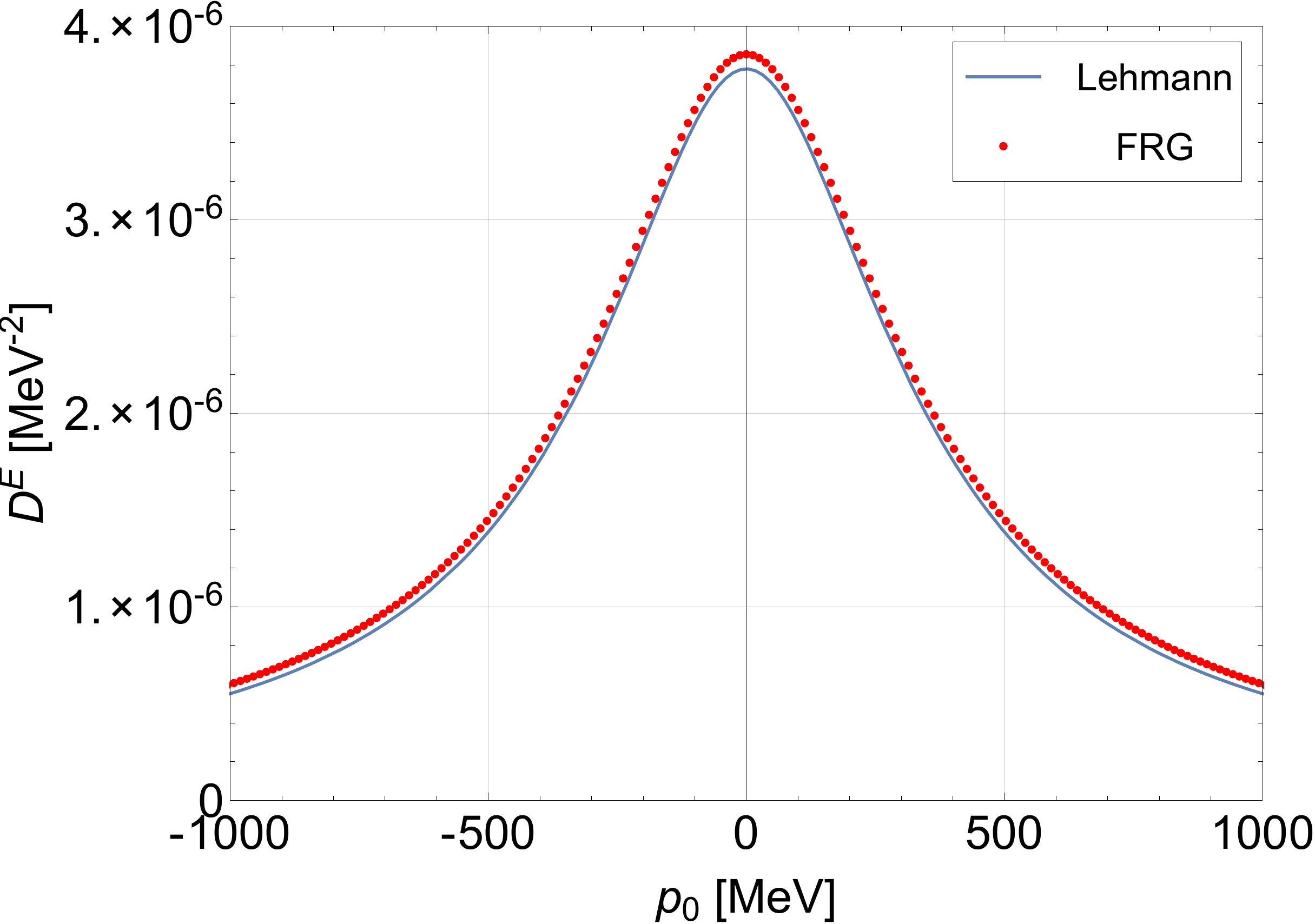}
	\includegraphics[width=0.49\textwidth]{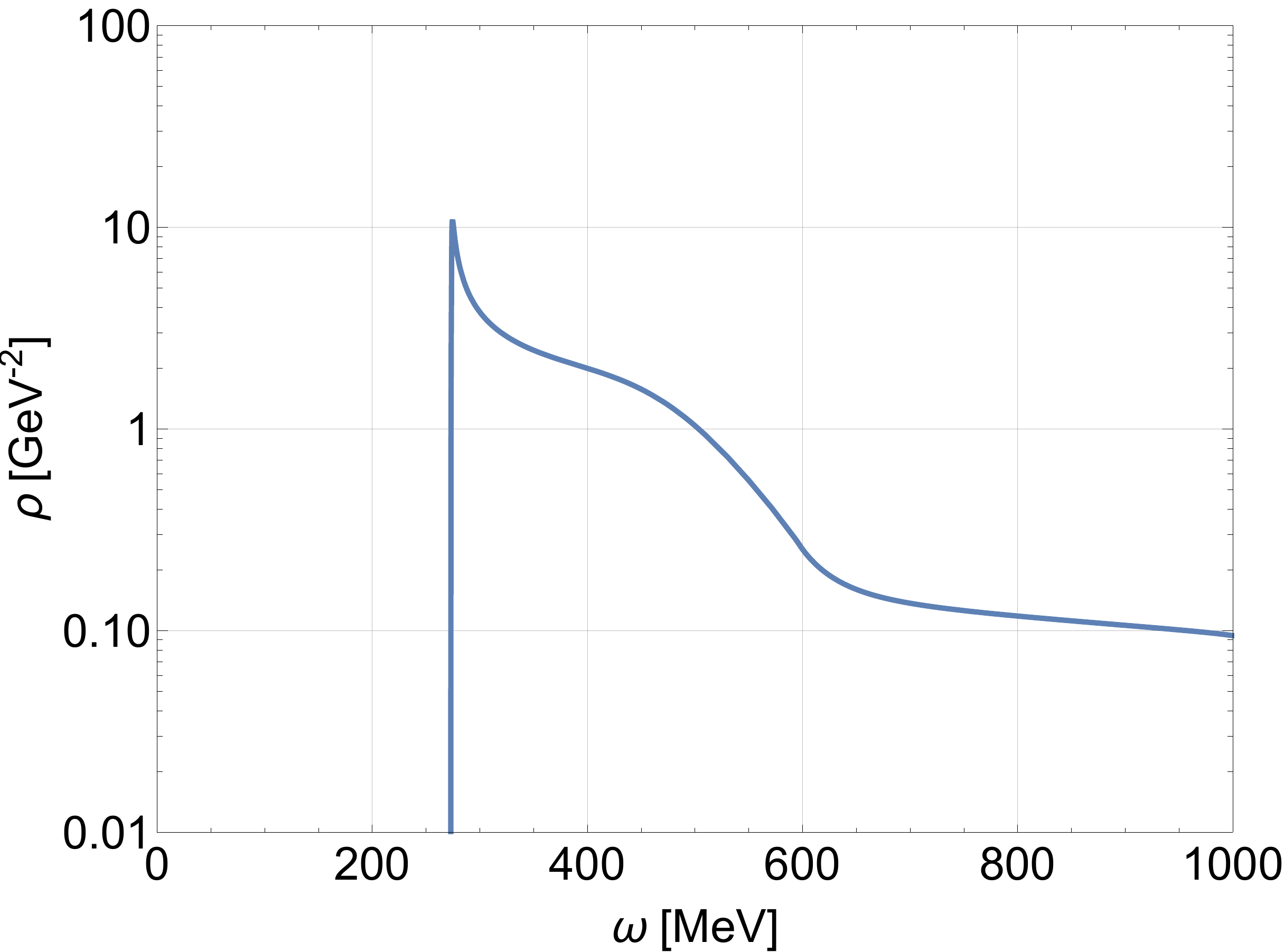}
	\caption{Left: The Euclidean sigma-meson propagator $D^E(p_0)$ is shown at the Matsubara energies $p_0=2\pi n T$ for $T=2$~MeV in comparison to the propagator obtained from the Lehmann representation, Eq.~(\ref{eq:Lehmann_p0}). Right: The corresponding input spectral function from the aFRG calculation.}
	\label{fig:Sigma_T2_Lehmann}
\end{figure}

\begin{figure}[b!]
	\centering\includegraphics[width=0.49\textwidth]{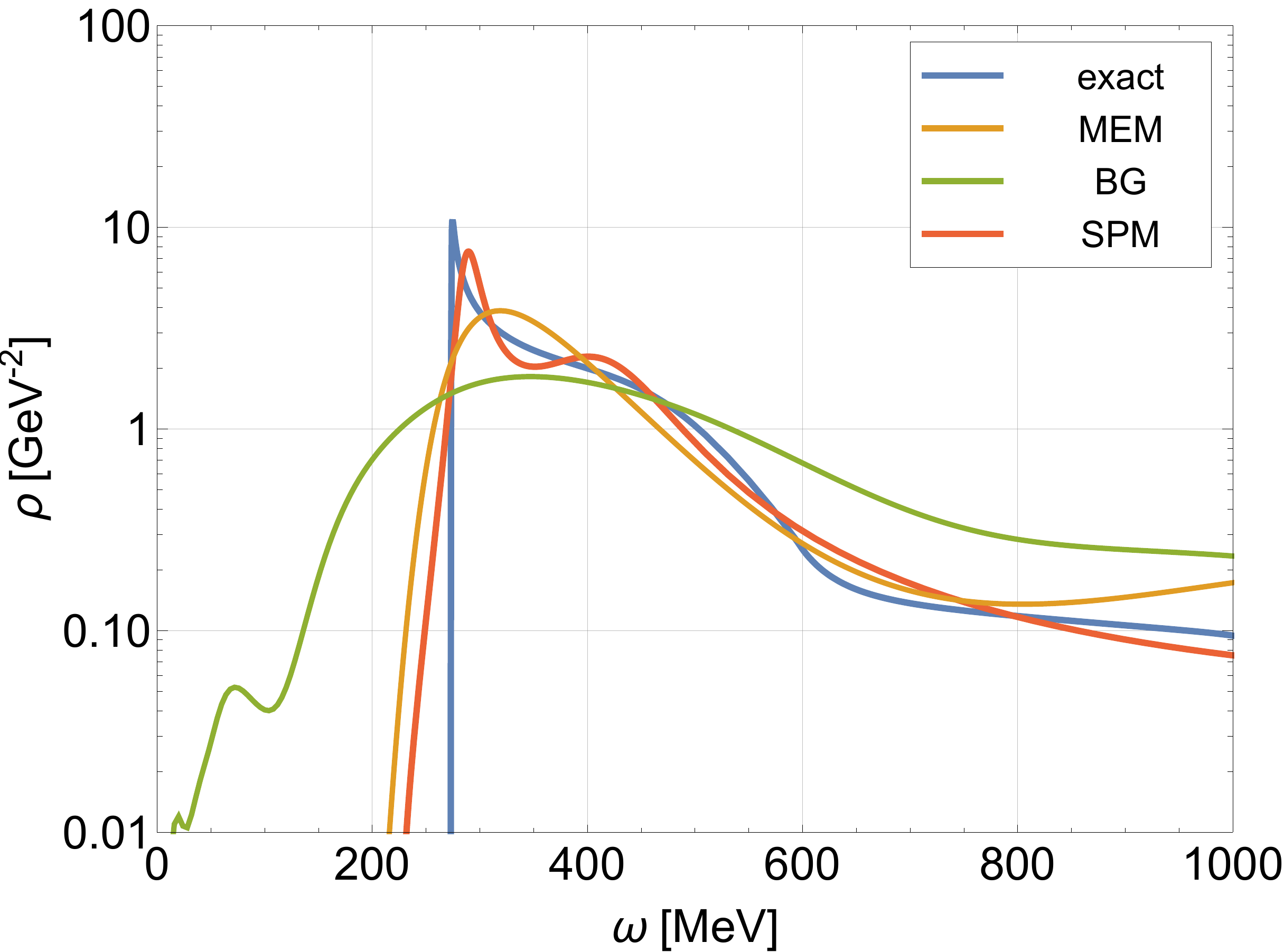}
	\caption{Comparison of the aFRG spectral function for the sigma meson at $T=2$~MeV with the reconstructed spectral functions obtained by using Euclidean data as input for the Maximum Entropy Method, the Backus-Gilbert method and the Schlessinger point method.}
	\label{fig:Sigma_T2_reconstruction}
\end{figure}

We now turn to the sigma spectral function at a temperature of $T=2$~MeV, see Fig.~\ref{fig:Sigma_T2_Lehmann}, which has a pronounced peak at the onset of the continuum at $\omega\approx 280$~MeV. Up to very small cutoff effects, the data points on $D^E(p_0)$ again agree well with the propagator obtained from the Lehmann representation. The reconstructed spectral functions obtained from MEM, the BG method and the SP method are shown in Fig.~\ref{fig:Sigma_T2_reconstruction}. The sharp near-threshold peak is best reconstructed by the SP method, followed by the MEM and the BG method. The continuum at higher energies is captured well by the SP method while the quality of the reconstruction is a bit worse for MEM while the BG method has difficulties to reproduce the structure and the magnitude of the continuum.

\subsection{Pion at $T=50$~MeV}

\begin{figure}[b!]
	\includegraphics[width=0.49\textwidth]{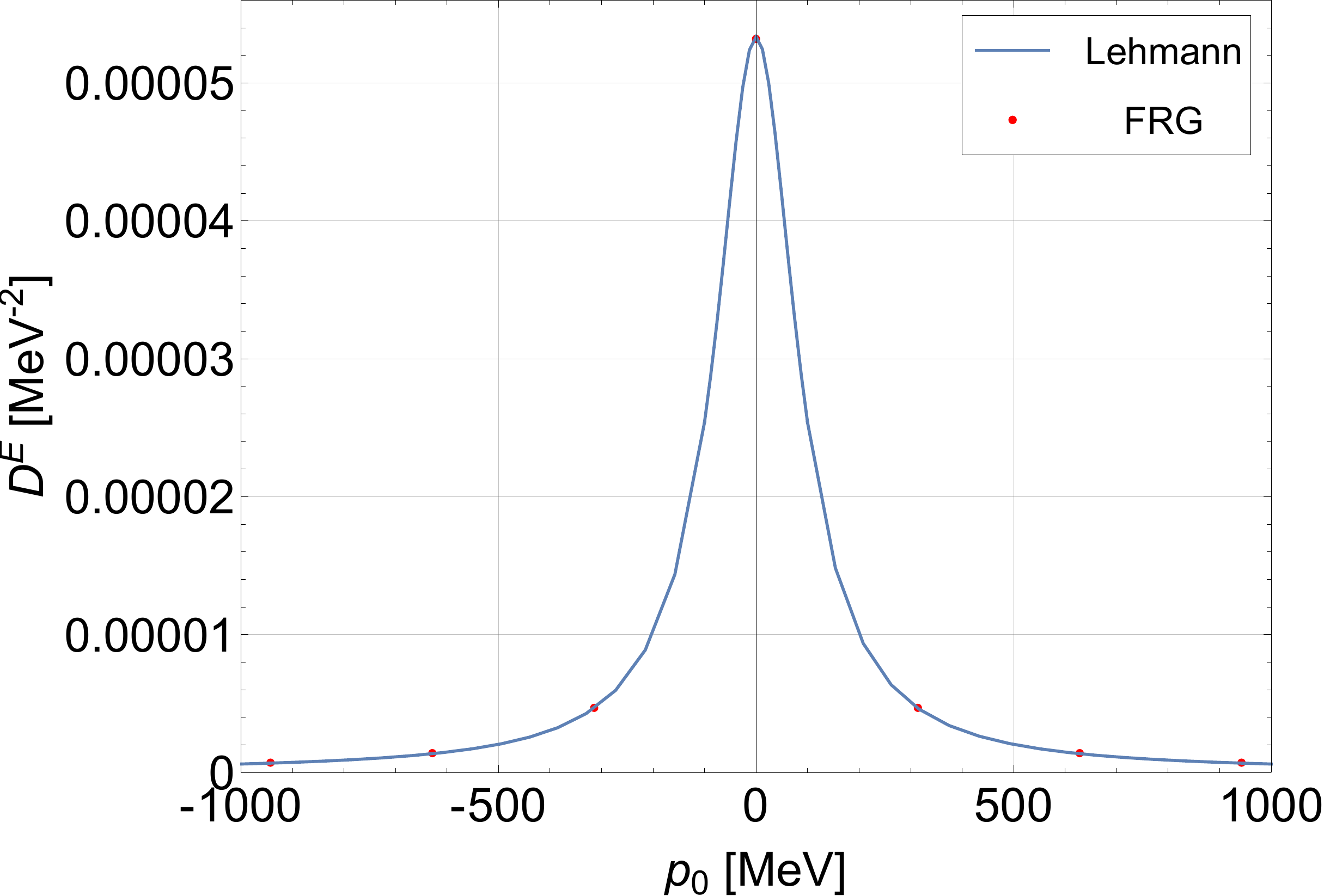}
	\includegraphics[width=0.49\textwidth]{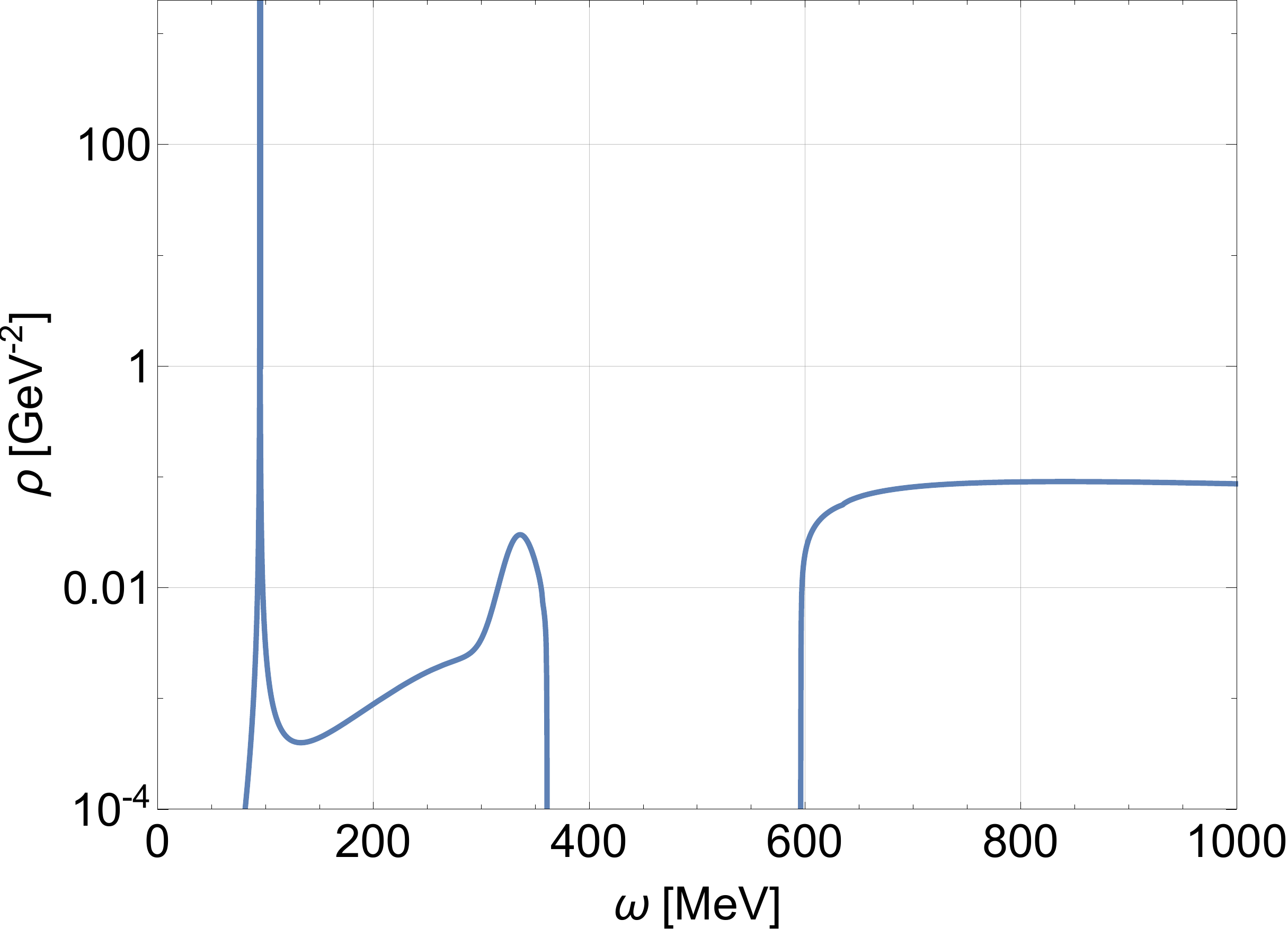}
	\caption{Left: The Euclidean pion propagator $D^E(p_0)$ is shown at the Matsubara energies $p_0=2\pi n T$ for $T=50$~MeV in comparison to the propagator obtained from the Lehmann representation, Eq.~(\ref{eq:Lehmann_p0}). Right: The corresponding input spectral function from the aFRG calculation.}
	\label{fig:Pion_T50_Lehmann}
\end{figure}

\begin{figure}[b!]
	\centering\includegraphics[width=0.49\textwidth]{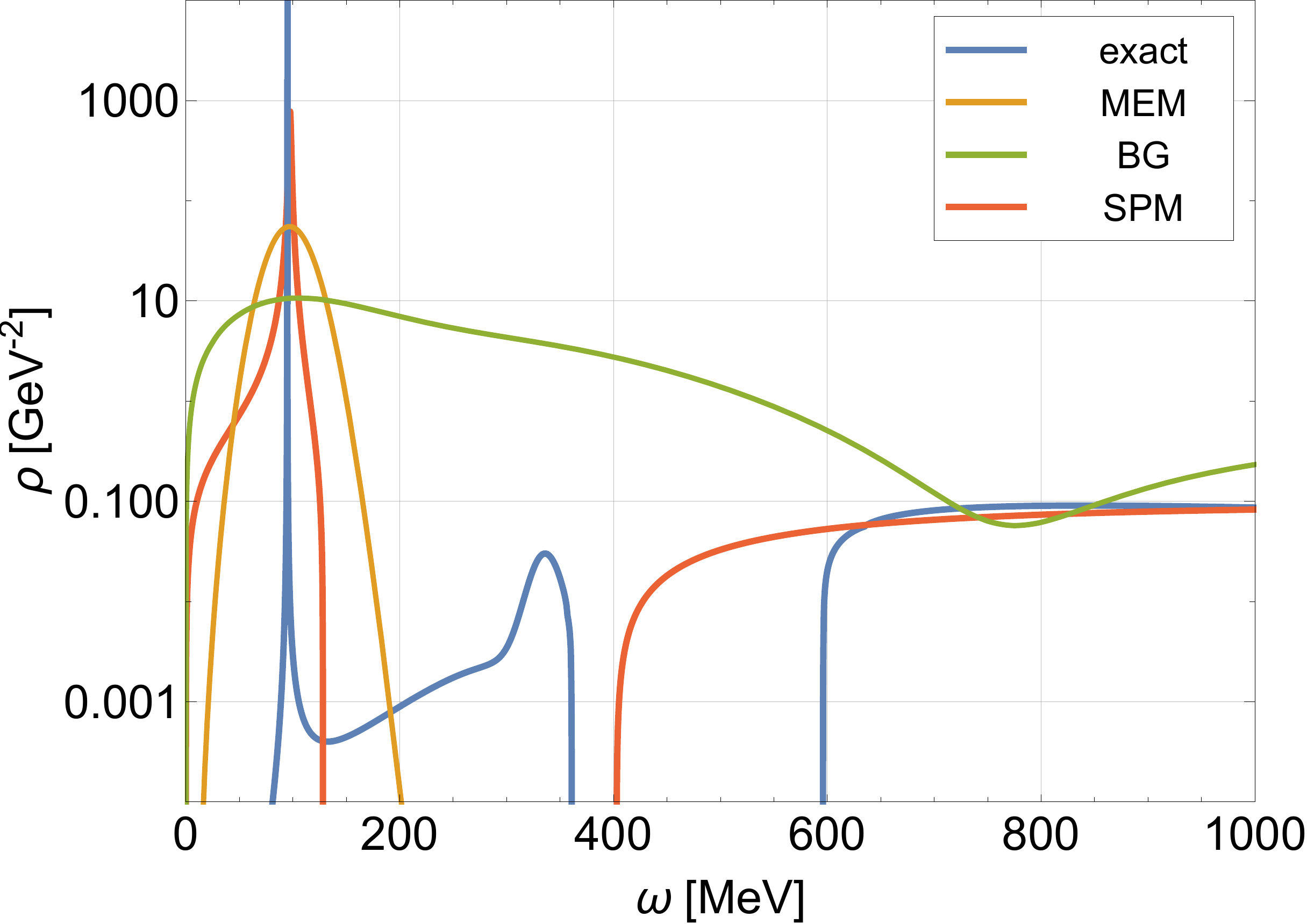}
	\caption{Comparison of the aFRG spectral function for the pion at $T=50$~MeV with the reconstructed spectral functions obtained by using Euclidean data as input for the Maximum Entropy Method, the Backus-Gilbert method and the Schlessinger point method.}
	\label{fig:Pion_T50_reconstruction}
\end{figure}

We will now study the pion spectral function at a temperature of $T=50$~MeV, see Fig.~\ref{fig:Pion_T50_Lehmann}, which has a sharp peak at $\omega\approx 95$~MeV embedded in a continuum ranging up to $\omega\approx 360$~MeV and a second continuum starting at $\omega\approx 600$~MeV. The physical origins of the various contributions to the spectral function, i.e.~the various decay and scattering processes in the heat bath, are explained in detail in \cite{Tripolt2014}.
Also with these rather complicated structures, the Euclidean propagator $D^E(p_0)$ obtained from the standard FRG calculation in thermal equilibrium and the propagator obtained from the Lehmann representation based on the aFRG solution agree very well as seen in Fig.~\ref{fig:Pion_T50_Lehmann}. This is a quite non-trivial consistency check of the underlying calculation schemes.

Due to the sizable temperature on the other hand, the separation between successive Matsubara modes, $\Delta p_0 \approx 314$~MeV, is already quite large compared to the relevant range of frequencies $\omega$ below 1~GeV as probed in the aFRG calculation of the spectral function. The discrete Euclidean $D^E(p_0)$ data obtained for $p_0 \gg 1 $~GeV, far outside this range, merely acts as an interpolation in the sampling of the Euclidean time propagator, $G^E(\tau)$, and does not contain any relevant information to constrain the contributions from frequencies below 1~GeV to its spectral representation in Eq.~(\ref{eq:Lehmann:tau}). The rapidly decreasing number of significant input points at increasing temperatures therefore generally makes reconstructions increasingly difficult. The reconstructed spectral functions obtained from MEM, the BG method and the SP method are shown in Fig.~\ref{fig:Pion_T50_reconstruction}. We find that the sharp peak is best reconstructed by the SP method, followed by MEM, while the BG method yields a very broad peak, yet at the correct position also in this case. The continuum is again best captured by the SP method while it can only be guessed from the BG method and is not seen by MEM at all.

\subsection{Sigma at $T=140$~MeV}

\begin{figure}[h!]
	\includegraphics[width=0.49\textwidth]{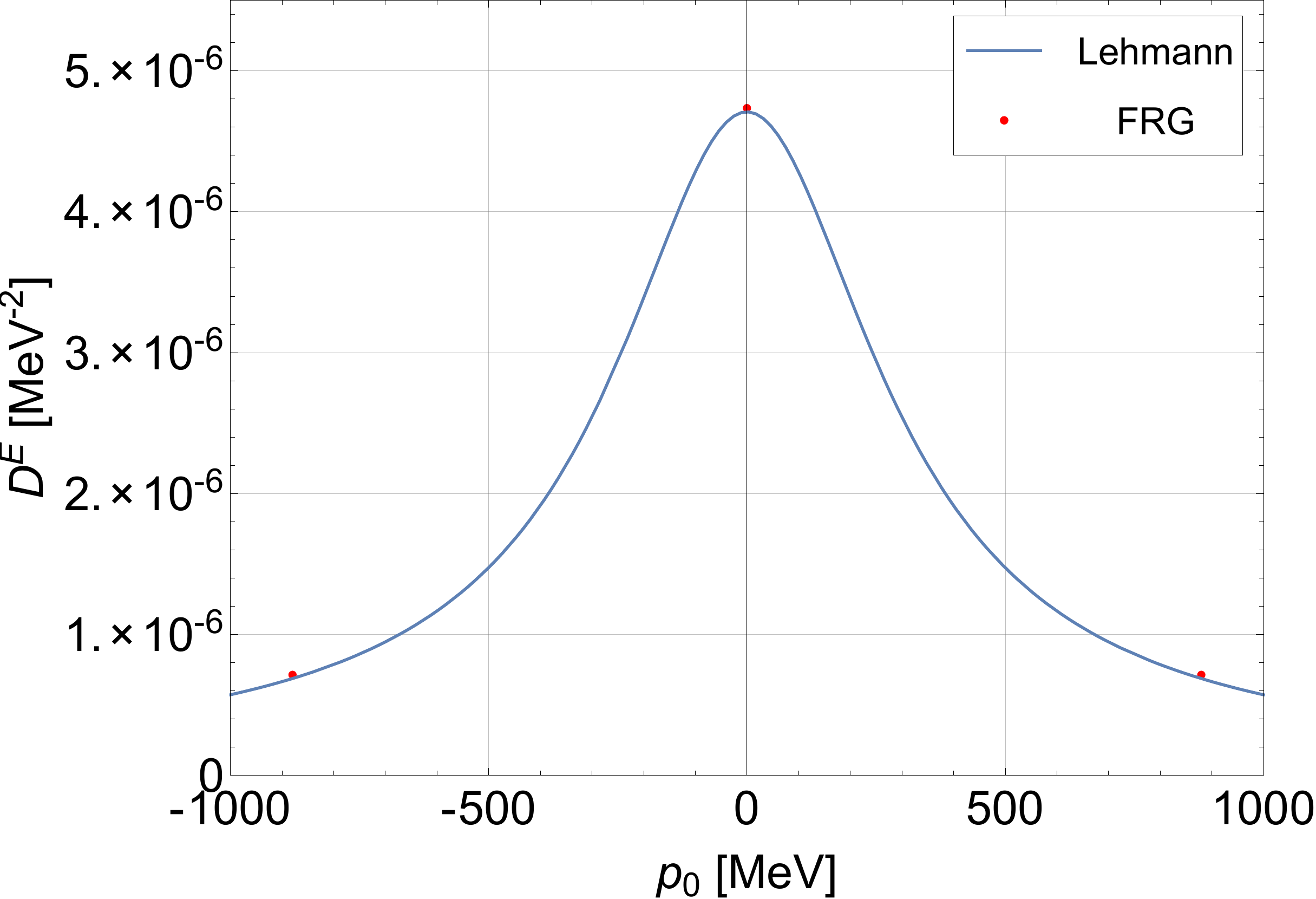}
	\includegraphics[width=0.49\textwidth]{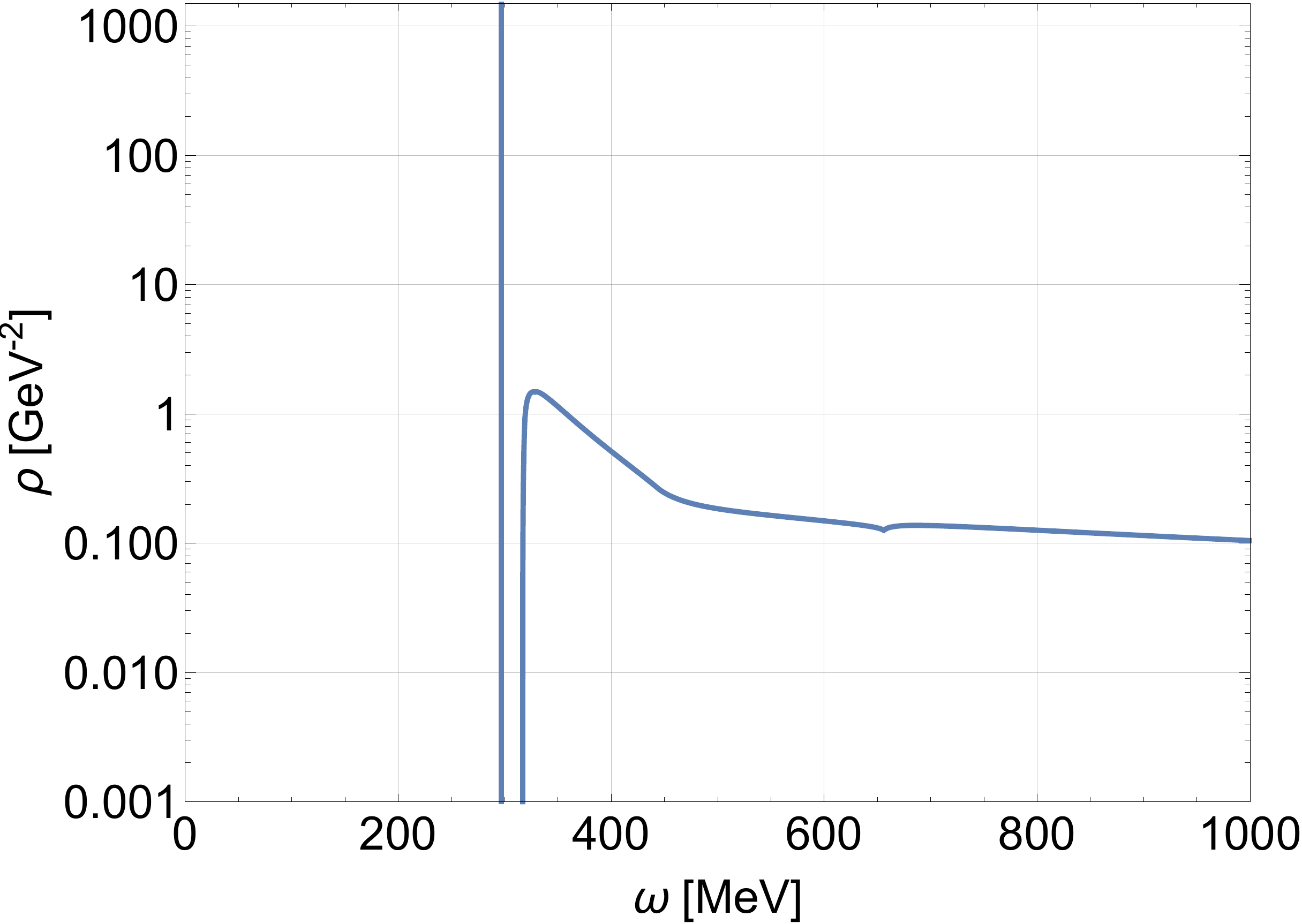}
	\caption{Left: The Euclidean sigma propagator $D^E(p_0)$ is shown at the Matsubara energies $p_0=2\pi n T$ for $T=140$~MeV in comparison to the propagator obtained from the Lehmann representation, Eq.~(\ref{eq:Lehmann_p0}). Right: The corresponding input spectral function from the aFRG calculation.}
	\label{fig:Sigma_T140_Lehmann}
\end{figure}

\begin{figure}[h!]
	\centering\includegraphics[width=0.49\textwidth]{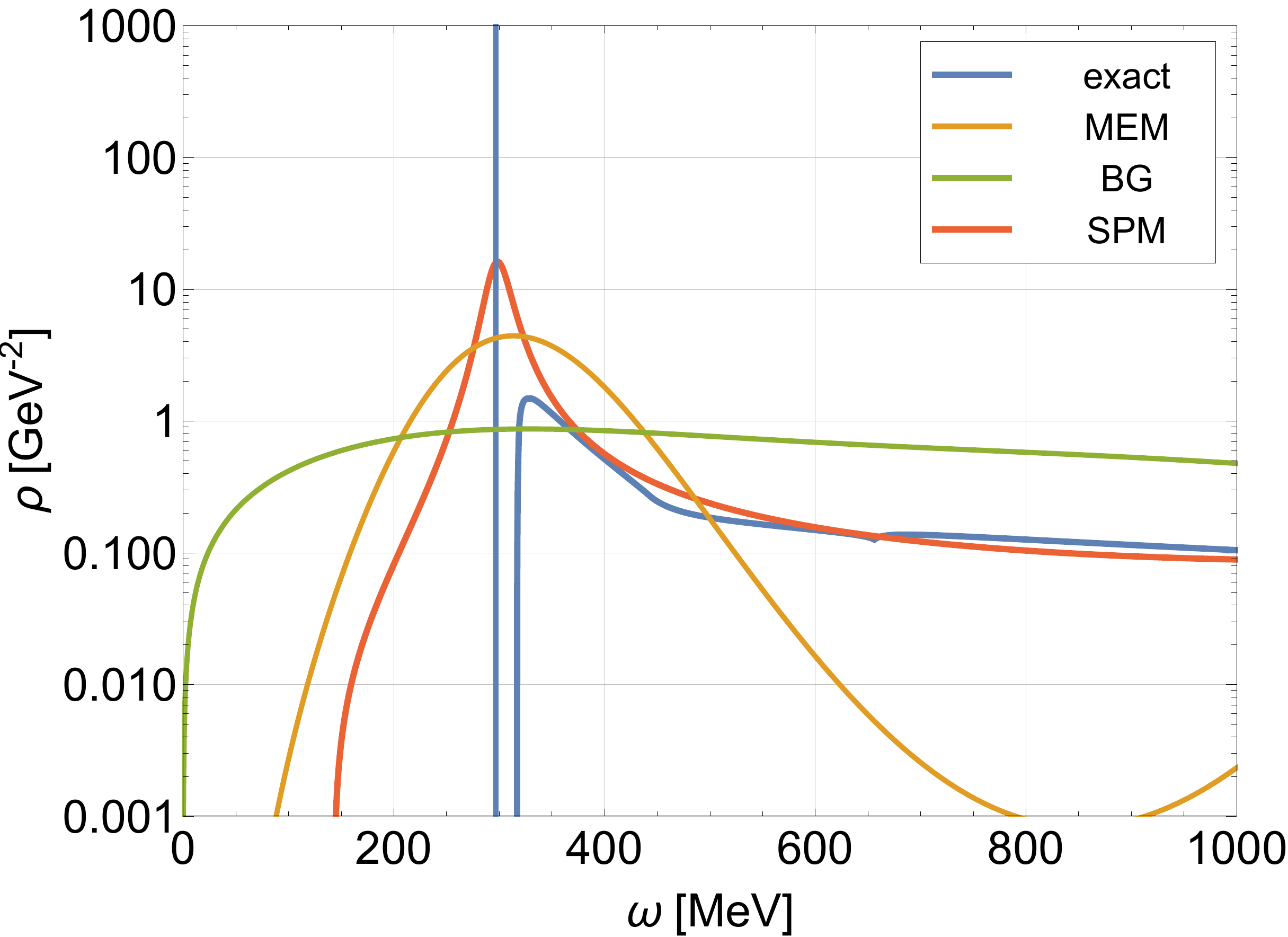}
	\caption{Comparison of the aFRG spectral function for the sigma meson at $T=140$~MeV with the reconstructed spectral functions obtained by using Euclidean data as input for the Maximum Entropy Method, the Backus-Gilbert method and the Schlessinger point method.}
	\label{fig:Sigma_T140_reconstruction}
\end{figure}

As a last example for FRG data we now turn to the sigma spectral function at a temperature of $T=140$~MeV, see Fig.~\ref{fig:Sigma_T140_Lehmann}, which has a pronounced cusp at the start of the continuum at $\omega\approx 280$~MeV. The agreement between the Euclidean propagator $D^E(p_0)$ obtained from the FRG calculation and the propagator obtained from the Lehmann representation is again very good. However, the separation of the Matsubara modes is already very large at $T=140$~MeV which makes the reconstruction of the spectral function very difficult for all three methods, see Fig.~\ref{fig:Sigma_T140_reconstruction}. The delta function is reproduced as a broad peak by the SP method and MEM while the BG method only shows an extremely broad maximum at roughly the correct energies. The continuum part is again best captured by the SP method while the other two methods do not give any reliable result.

\section{Lattice QCD data}
\label{sec:Lattice}

In order to compare the three methods on data with realistic sampling and errors as obtained from present-day lattice Monte-Carlo simulations, we will now apply them to the Euclidean-time correlator in the isovector vector-meson channel from the lattice QCD simulations with two flavors of clover-improved Wilson quarks of \cite{BrandtFrancisJaegerEtAl2016} as a physical example. In particular, we use the data for the vacuum correlator in their Table VII from the  $64^3 \times 128$ lattice with a lattice spacing of $a \simeq 0.0486$~fm. From  $T=1/(a N_\tau)$ with $N_\tau = 128$ this corresponds to a temperature of $T \simeq 32$~MeV. The pion mass is stated to be $m_{\pi} \simeq 270$~MeV and hence somewhat heavier than the physical one.  The spatial lattice size $L = a N_s $ with $N_s =64 $ is just above three Fermi and hence sufficiently much larger than the Compton wavelength of the pion, $m_\pi L \simeq 4.2$. For further details we refer to the original paper, especially Tables III and VII in \cite{BrandtFrancisJaegerEtAl2016}. One expects the vector-meson spectral function to contain a peak at the mass $m_\rho$ of the $\rho$-meson together with a continuum for $\omega \ge \Omega_0$ starting at some threshold $\Omega_0 > m_\rho$. In \cite{BrandtFrancisJaegerEtAl2016}, a fit model was therefore used of a form approaching the perturbative expectation at large $\omega$, 
\begin{align}
\label{eq:rhofit}
  \rho(\omega) = a_V \, \delta(\omega - m_\rho) + \frac{3\kappa_0 }{4\pi^2}
  \theta(\omega - \Omega_0) \, \omega^2\tanh\big(\beta\omega/4\big) \,  . 
\end{align}
The relevant fit parameters, in lattice units, were reported as $a m_\rho = 0.205(5) $ and $a\Omega_0 = 0.319(16)$  \cite{BrandtFrancisJaegerEtAl2016}, corresponding to a $\rho$-meson mass of $m_\rho \approx 830 $~MeV and a threshold at $\Omega_0 \approx 1300 $~MeV.

\begin{figure}[t]
	\centering\includegraphics[width=0.49\textwidth]{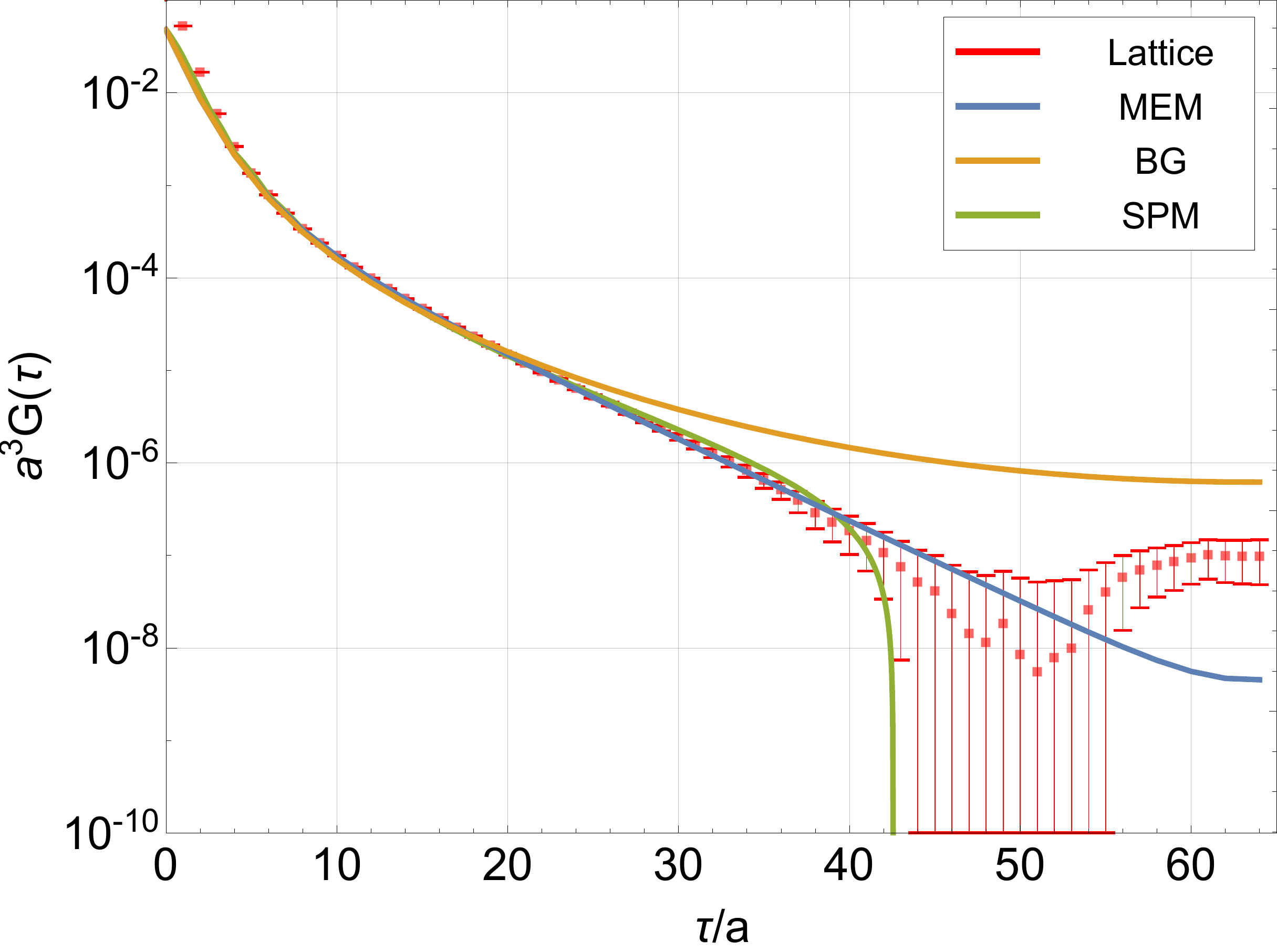}
	\centering\includegraphics[width=0.49\textwidth]{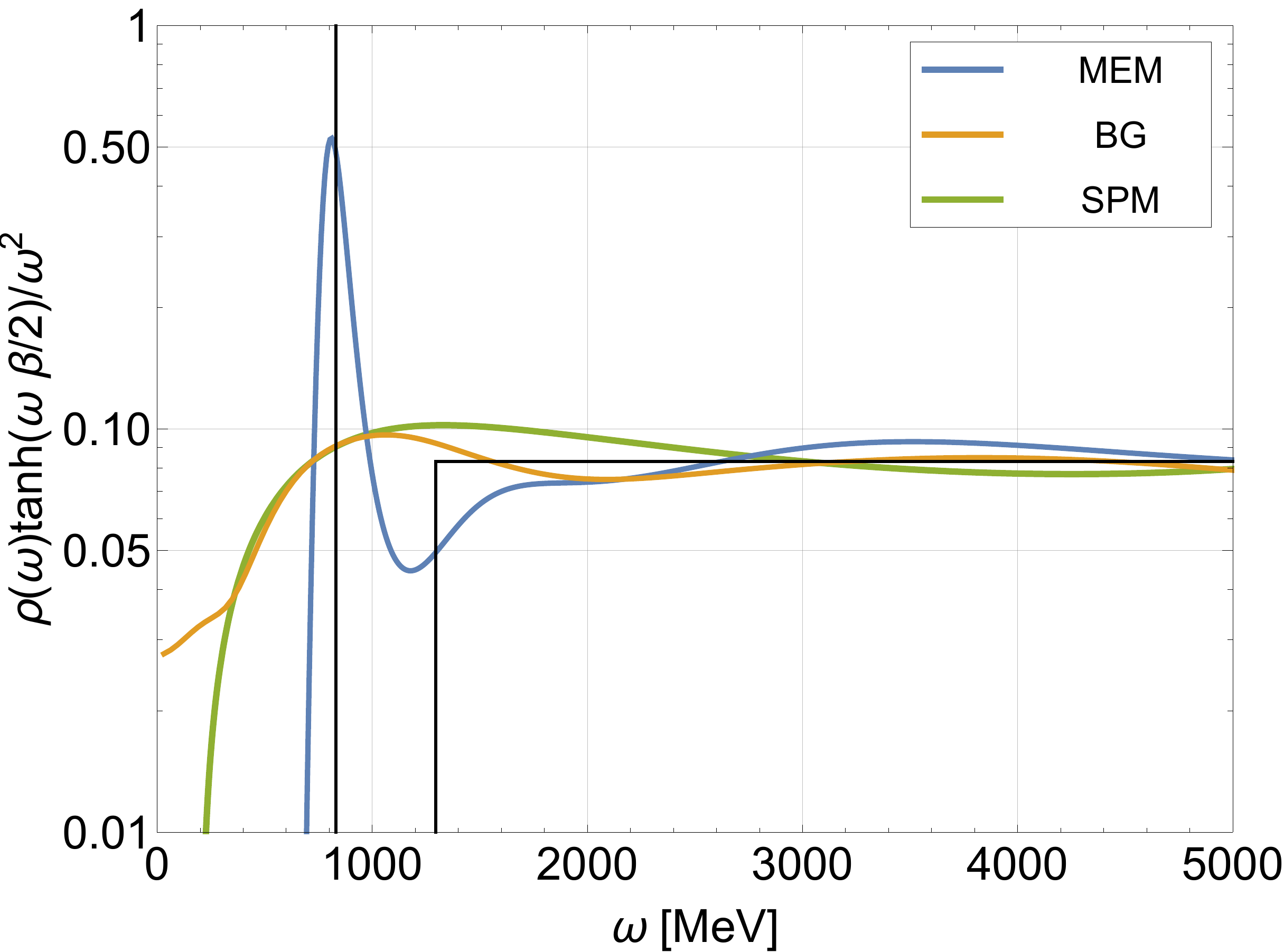}	
	\caption{Data from \cite{BrandtFrancisJaegerEtAl2016} on the isovector current correlator in two-flavor lattice QCD with pion mass $m_{\pi} \simeq 270$ MeV at temperature $T = 32$ MeV (left) compared to the Lehmann representations (\ref{eq:Lehmann:tau}), as obtained from the three reconstructed spectral functions shown on the right: via MEM, the BG method and the SP method (together with the model fit from \cite{BrandtFrancisJaegerEtAl2016}).}
	\label{fig:lattice_data}
\end{figure}

\begin{figure}[t]
	\centering\includegraphics[width=0.49\textwidth]{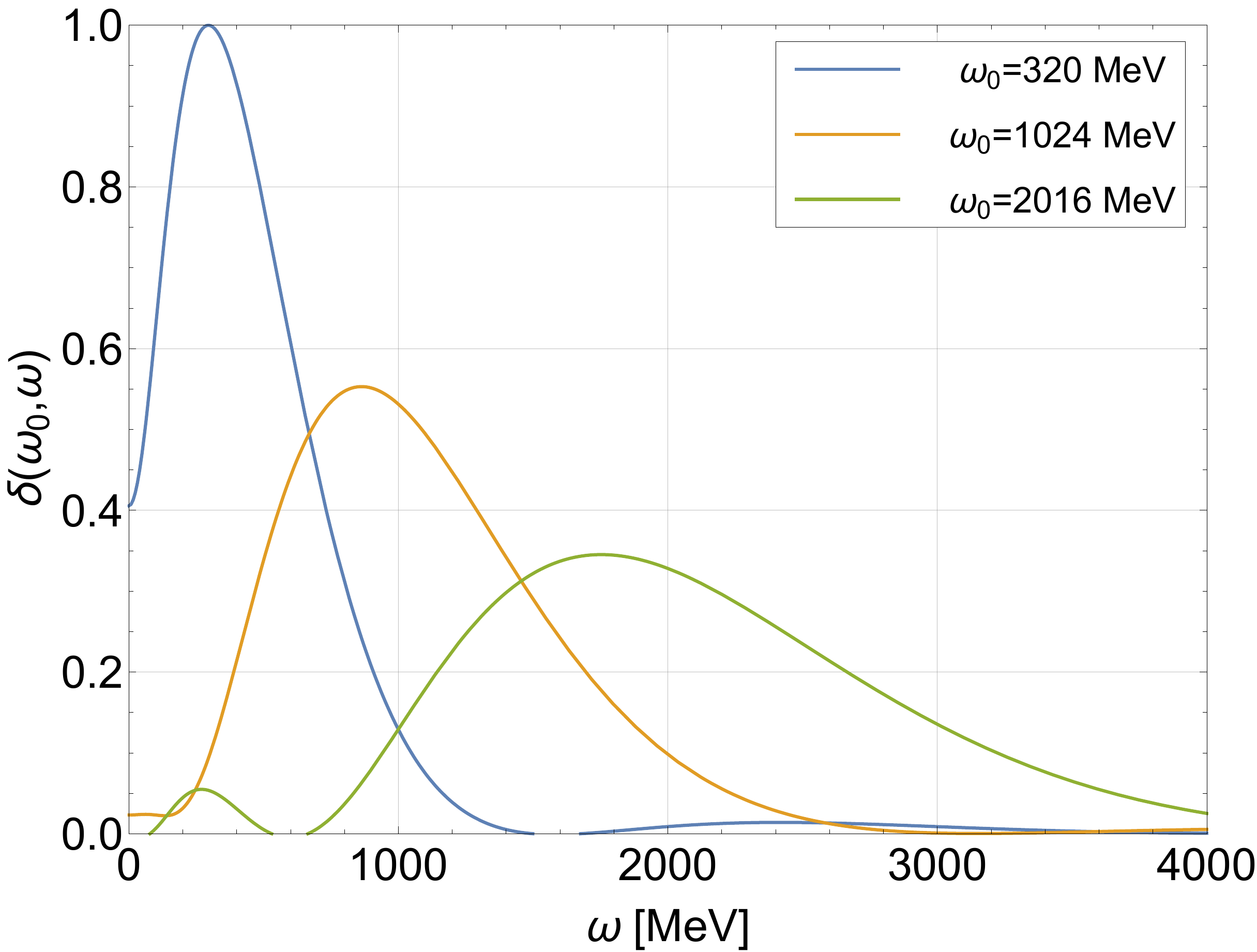}
	\caption{Resolution functions $\delta(\omega_0, \omega)$ for different $\omega_0$ for the reconstruction shown in Fig.~\ref{fig:lattice_data}. Resolution functions are rescaled such that the one for the smallest $\omega_0$ has its peak value at 1.}
	\label{fig:resolution_functions}
\end{figure}

The lattice data for the isovector current correlator in the Euclidean time (at zero momentum), in lattice units $a^3 G(\tau)$, is shown in the left panel of Fig.\,\ref{fig:lattice_data}, cf.~Fig.~1 in \cite{BrandtFrancisJaegerEtAl2016}. As it is symmetric around $\beta/2$, only the first 64 time slices provide independent information. One observes that the relative error of the data increases with increasing $\tau$, reaching a maximum at around $\tau/a \simeq 50$. Moreover, in this large-time region there appears to be a positivity issue because the central values of $G(\tau)$  are not convex to the above as they ought to be for a positive spectral function, for which
\begin{equation} 
\frac{d^2}{d\tau^2}  \ln G(\tau) \ge \langle \omega^2 \rangle_\tau -  \langle \omega \rangle^2_\tau \ge 0\, ,
\end{equation}
because $\langle \omega \rangle_\tau $ and $  \langle \omega^2 \rangle_\tau $ are the moments of the positive measure given by the integrand in Eq.~(\ref{eq:Lehmann:tau}) at every time $\tau \in [0,\,\beta] $. Of course within the errors this violation might not be a very significant one, but it is an additional challenge for the reconstruction methods. In the MEM reconstruction, which was used also in \cite{GublerSatow2017}, we therefore only use the first 48 time steps here and furthermore omit the lowest four data points which are most affected by the lattice discretization. In the SP method we first have to apply the discrete Fourier transform (\ref{eq:FT}). Without changing the temperature it is then not possible to omit data points from the analysis without additional model bias.
 We have tested, however, that setting the non-convex data points in $G(\tau)$ for $\tau/a\geq 48$ to zero did not lead to any significant changes. On one hand, these values are simply too small (by several orders of magnitude) to have any substantial impact, and on the other, the corresponding correlation function would of course not be convex to the above either. We therefore decided not to introduce such an additional model bias, neither in the SP nor the BG method.

The extracted spectral functions of the three methods are given on the right-hand side of Fig.\,\ref{fig:lattice_data}. 
MEM extracts a clear peak at around $a\omega = 0.2$ corresponding to $m_\rho \approx 820 $~MeV, which is somewhat lower but still consistent with the analysis of \cite{BrandtFrancisJaegerEtAl2016} within the errors, where the data was fitted to the form in Eq.~(\ref{eq:rhofit}) describing the stable particle pole at $m_\rho$  and the  continuum starting at $\Omega_0$. The BG method also finds a bump in the same energy region, while for the  SP method the peak is least visible, but a maximum is seen at about $a\omega = 0.3$. All three methods are able to reconstruct the continuum at larger energies, consistent with the expectations from perturbative QCD. 

The left-hand side of Fig.~\ref{fig:lattice_data} also shows the correlation functions as obtained by inserting the reconstructed spectral functions back into the Lehmann representation. MEM gives a convex correlator in good agreement with the input data while the SPM correlator agrees well for $\tau/a<40$ but then becomes negative. The BG method also produces a convex correlator with a good agreement with the lattice data up to $\tau/a\approx 20$.

We note that the BG method does not produce the spectral function itself but only its convolution with the resolution function, i.e.~a smeared-out version, see also Sec.~\ref{sec:BG}. Examples of the resolution function for this particular reconstruction are plotted in Fig.~\ref{fig:resolution_functions}. First of all, the resolution functions can be non-symmetrical. This means that the position of the peak of the resolution function generally does not coincide with its formal center $\omega_0$. Thus there is an ambiguity as to which frequency should be formally used in the  Lehmann representation. We checked, however, that the error introduced by this ambiguity is negligible. Another unfortunate feature of the resolution functions is that their widths generally increase with increasing $\omega_0$. Thus the BG method is able to  reconstruct sharp features of spectral functions only at lower frequencies.  At large frequencies the spectral function becomes more and more smoothed out.

In all, there are quite large differences between the outcomes of the three reconstruction methods at the present level of accuracy and sampling. 
In agreement with the reliability estimates in Fig.~\ref{fig:regime_applicability} we observe that the lattice data here is right at the boundary of the applicability regions for BG and MEM, while it lies outside that for the SP method. Therefore, we conclude that even higher accuracy and sampling rates will be necessary for more precise and unambiguous determinations of spectral functions in the future.

\section{Electrical Conductivity of Graphene}
\label{sec:conductivity}

In this section we study an example from condensed matter physics, namely the electrical conductivity of graphene. Quantum Monte Carlo (QMC) simulations based on a tight-binding model are a typical \textit{ab-initio} tool used for the calculation of the electrical conductivity of graphene \cite{conductivity1, conductivity2}. Since QMC provides only Euclidean-time data, the analytic continuation is an inevitable step in these calculations. In the following we will study the simpler case of a free tight-binding model for graphene, without the need for QMC simulations, and test the applicability of the three analytic continuation methods discussed in this work to extract the static conductivity of Dirac quasiparticles in 2+1 dimensions from the Euclidean time data.

The starting point is the tight-binding Hamiltonian on graphene's hexagonal lattice, which describes the hopping of electrons between the nearest-neighbor lattice sites: 
\begin{eqnarray}
\label{tbHam}
\hat{H}= - \kappa \sum_{<x,y>} \left( { \hat{a}^{\dag}_{y, \uparrow} \hat{a}_{x, \uparrow} +  \hat{a}^{\dag}_{y, \downarrow} \hat{a}_{x, \downarrow} + h.c.} \right).
\end{eqnarray}
Here $\kappa = 2.7$~eV denotes the hopping amplitude, $\hat{a}^{\dag}_{x, \uparrow}$, $\hat{a}_{x, \uparrow}$ and $\hat{a}^{\dag}_{x, \downarrow}$, $\hat{a}_{x, \downarrow}$ are creation/annihilation operators for spin up and spin down electrons situated at the current lattice site. The spatial index $x=\{s, \xi\}$ consists of a sublattice index $s=1,2$ and the two-dimensional coordinate $\xi=\{\xi_1, \xi_2\}$ of the unit cell in a rhombic lattice (the rhombic lattice with two atoms in each cell forms a hexagonal lattice structure). The sum is performed over all pairs of nearest neighbors. In real calculations we deal with a finite sample with periodic boundary conditions in both spatial directions.  This Hamiltonian hosts free Dirac quasiparticles at sufficiently low energies close to the Fermi level at half filling. 

The frequency-dependent conductivity $\sigma(\omega)$ is calculated from the following Green-Kubo relation:
\begin{equation}
G_{JJ}(\tau) = \int_0^\infty  \sigma(\omega) \frac {\omega \cosh (\omega(\beta/2 -\tau))} {\pi \sinh(\omega \beta/2)}  d\omega,
\label{GK_cond}
\end{equation}
where $G_{JJ}(\tau)$ is the current-current correlator 
\begin{equation}
G_{JJ}(\tau) =  \frac {T_{bc}} {3 \sqrt 3 N_s } \sum_{\xi}  \frac {\mbox{Tr}  \left( { e^{-\beta \hat H} \hat J_{b}(\xi) e^{-\tau \hat H} \hat J_{c}(\xi) e^{\tau \hat H} }\right)  }{\mbox{Tr} \, e^{-\beta \hat H}}
\label{JJ_corr}
\end{equation}
calculated for the free fermions \ref{tbHam}  on the lattice  with $N_s$ unit cells with inverse temperature equal to $\beta$. We imply the summation over repeated indexes in the last formula.  $\hat J_{b}(\xi)$ is the operator of electromagnetic current flowing  from the site in the first sublattice with coordinate $x=\{1,\xi\}$  towards one of its nearest neighbors. There are three nearest neighbors and 3 possible directions of the current: $b=1,2,3$. The matrix $T_{bc}$ is defined as follows:
\begin{equation}
T_{bc}=d^2 \left\{ {  {1, \quad b=c}  \atop  {-1/2, \quad b \neq c}  }  \right.  ,
\label{matrix_T}
\end{equation}
where $d=0.124\, \mbox{nm}$ is the distance between nearest neighbors in the graphene lattice.
The explicit form of the electromagnetic current operator can be written as
\begin{equation}
\hat J_b (\xi) =  \sum_{\sigma=\uparrow, \downarrow} \hat J_{\sigma, b} (\xi) = i \kappa \sum_{\sigma=\uparrow, \downarrow}  \hat a^{\dag}_{\sigma, x} \hat a_{\sigma,y} + h.c., 
\end{equation}
where $x=\{1, \xi \}$, $y=\{2, \xi+\rho_b \}$ and shifts in the rhombic lattice are defined as follows: $\rho_1=\{0,0\}$, $\rho_2=\{-1,1\}$, $\rho_3=\{-1,0\}$. 
The final form of the free current-current correlator is written as the following combination of the free lattice fermionic propagators $\Gamma(x,\tau, x,\tau')$ (see \cite{conductivity_calc} for the derivation):
\begin{eqnarray}
G(\tau) = - \frac {2 T_{bc}} {3 \sqrt 3 N_s}   \sum_{x_1,x_2,y_1,y_2}\bigl [ j_b(y_1,x_1) \Gamma(x_1,0, x_2,\tau) j_c(x_2, y_2) \Gamma (y_2,\tau,y_1, 0) \bigr ]  +
\nonumber \\
   \frac {4 T_{bc}} {3 \sqrt 3 N_s}  \; \sum_{x,y} \bigl [ j_b(y, x) \Gamma(x,0, y, 0) \bigr ]
\cdot  \sum_{x,y} \bigl [ j_c(y,x) \Gamma(x,\tau, y, \tau) \bigr ].
\label{lattice_correlation}
\end{eqnarray}
Here $j_b$ is current vertex:
\begin{equation}
\sum_{\xi} \hat J_{\sigma, b} (\xi) = \sum_{ x=\{s, \xi \} \atop y=\{s', \xi' \}}  \hat a^{\dag}_{\sigma, x} \hat a_{\sigma, y}  j_b (x,y).
\label{j_matrix}
\end{equation}
All calculations were done for the free current-current correlator in the tight-binding model of Eq.~(\ref{tbHam}) on a $48\times48$ lattice with a temperature of $1/\beta=0.125$~eV. The number of Euclidean time slices is $N_t=80$. 

The static conductivity of 2+1 dimensional Dirac quasiparticles usually is the quantity of most central interest to be extracted from this correlator. It is in principle given by the limit $\sigma(\omega)|_{\omega\rightarrow 0}$. However, the complicated structure of the spectral function makes a clear reconstruction difficult. The first problem is the Drude peak proportional to $\delta(\omega)$ \cite{Katsnelson2006} which appears in the spectral function calculated for any sample without open boundaries. Another difficulty is the drop-down of the spectral function at small but nonzero frequencies which reflects the finite sample size. Finally, at large frequencies $\omega\approx 2 \kappa$, we have a peak associated with the van Hove singularity in the density of states. All these features can be observed in the exact calculation of the full profile  $\sigma(\omega)$ presented in \cite{conductivity_calc} for the free tight-binding model (\ref{tbHam}). From this calculation on can see explicitly that the true static conductivity of the Dirac quasiparticles is represented by a plateau located in between the low frequency ``gap'' due to the finite lattice size and the high-frequency peak from the van Hove singularity. The aim of the analytic continuation methods in this case is therefore not to identify the position of the peaks, but to reproduce the correct value of the spectral function at the plateau and, of course, to reproduce the plateau itself. 

\begin{figure}[t]
	\centering\includegraphics[width=0.49\textwidth]{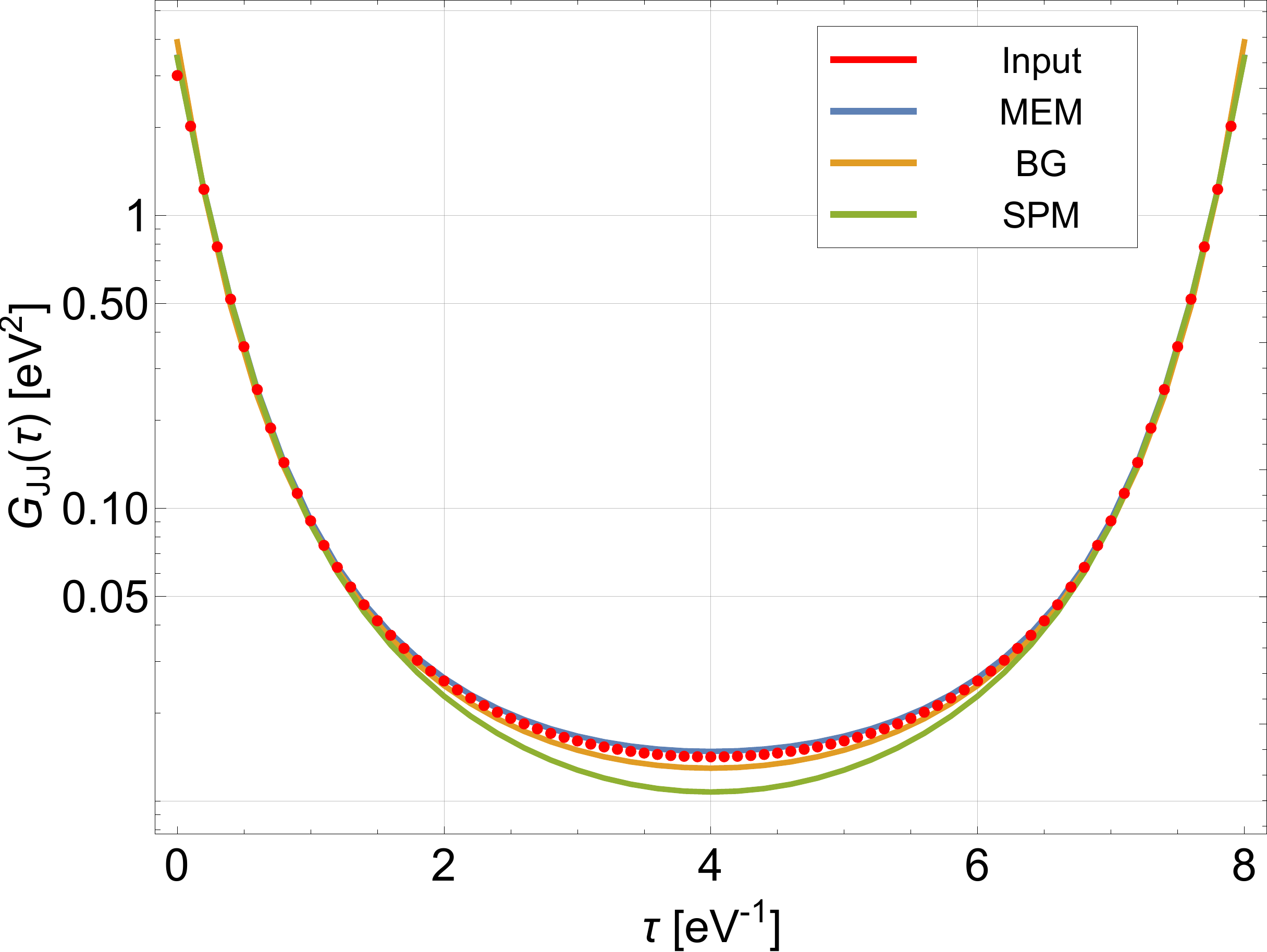}	
	\centering\includegraphics[width=0.49\textwidth]{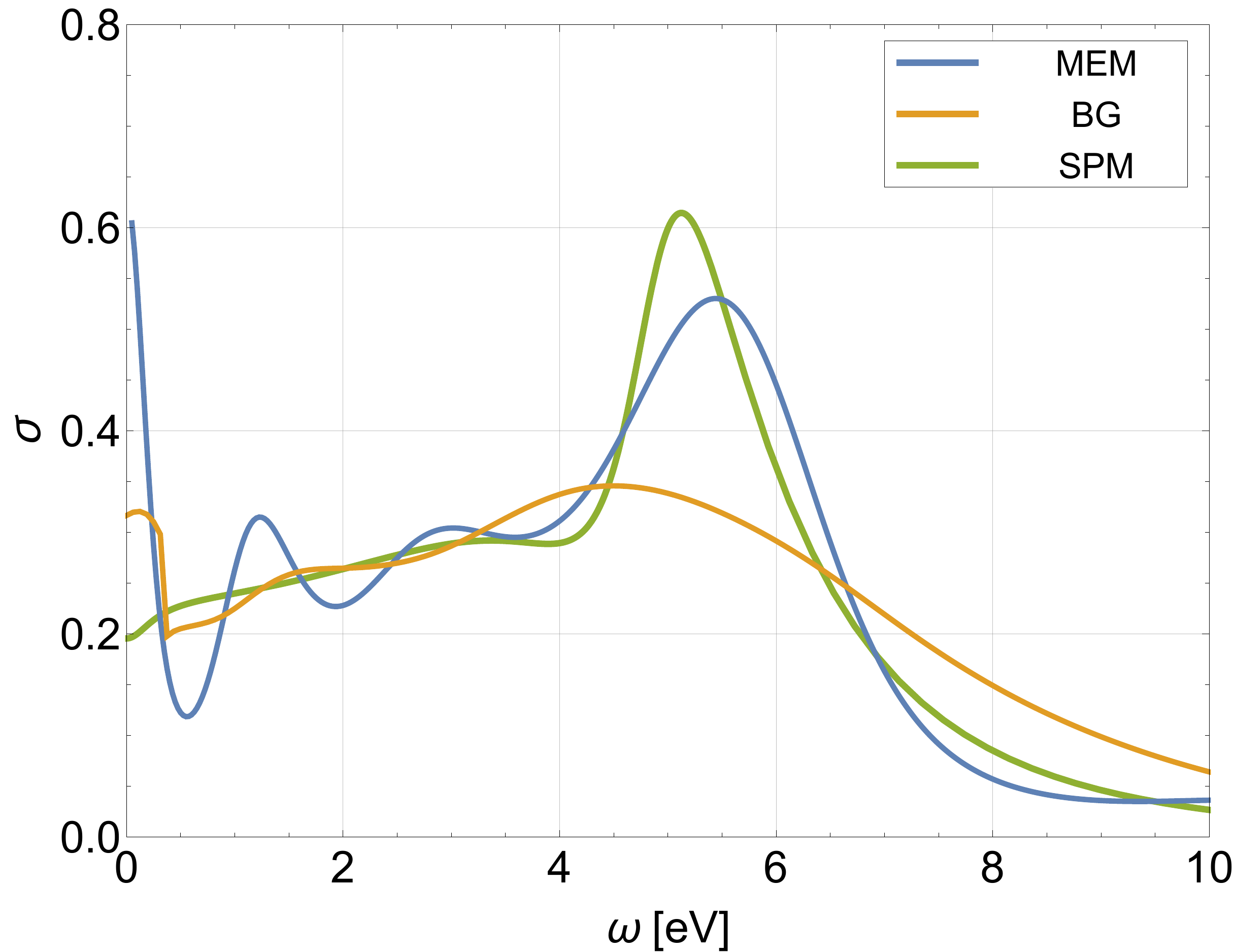}	
	\caption{Left: Input data on the current-current correlator in the tight-binding model for graphene compared to the Green-Kubo representations of Eq.~(\ref{GK_cond}), as obtained from the three reconstructed electrical conductivity functions shown on the right: via MEM, the BG method and the SP method.}
	\label{fig:conductivity}
\end{figure}

\begin{figure}[h!]
	\centering\includegraphics[width=0.49\textwidth]{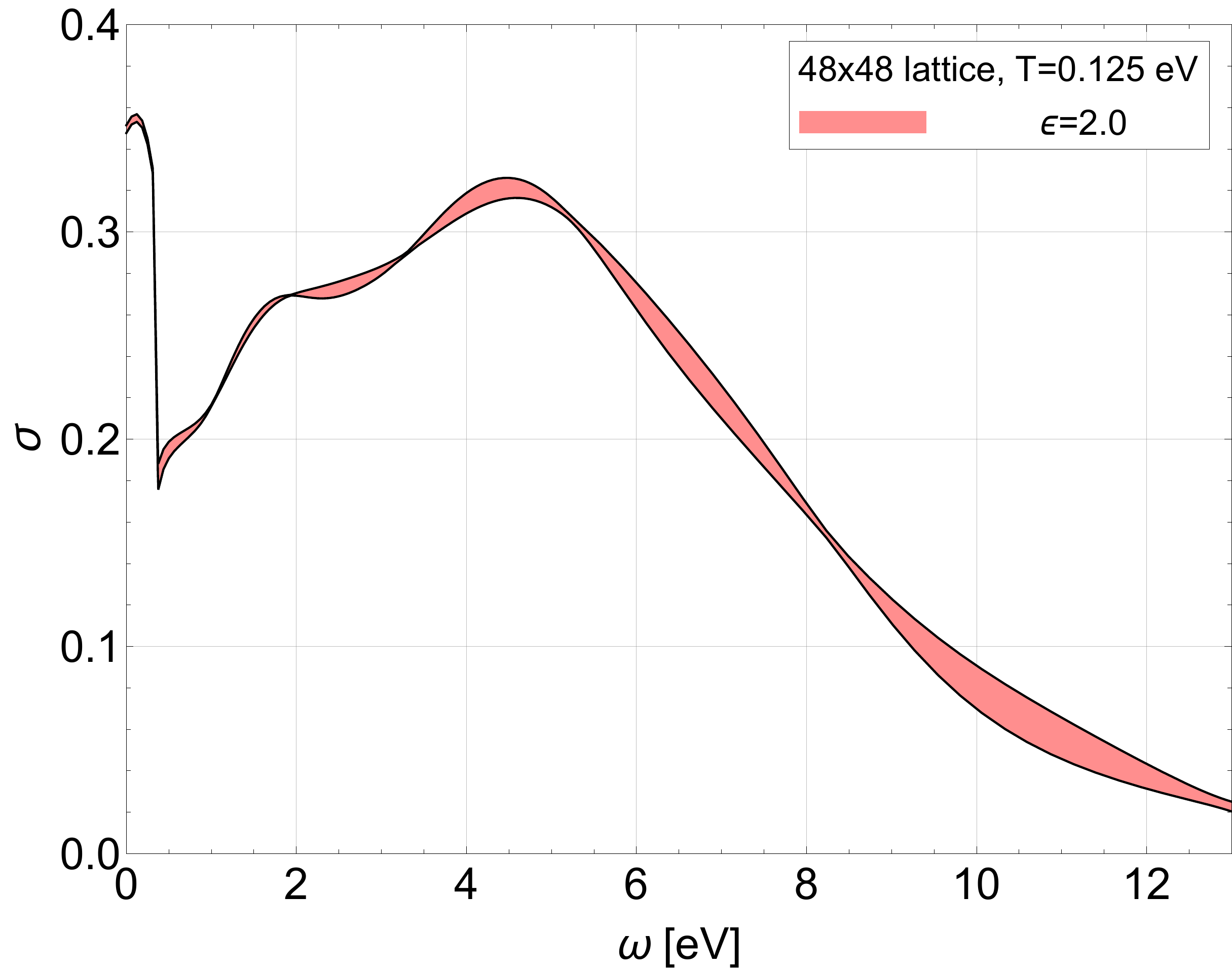}	
	\centering\includegraphics[width=0.49\textwidth]{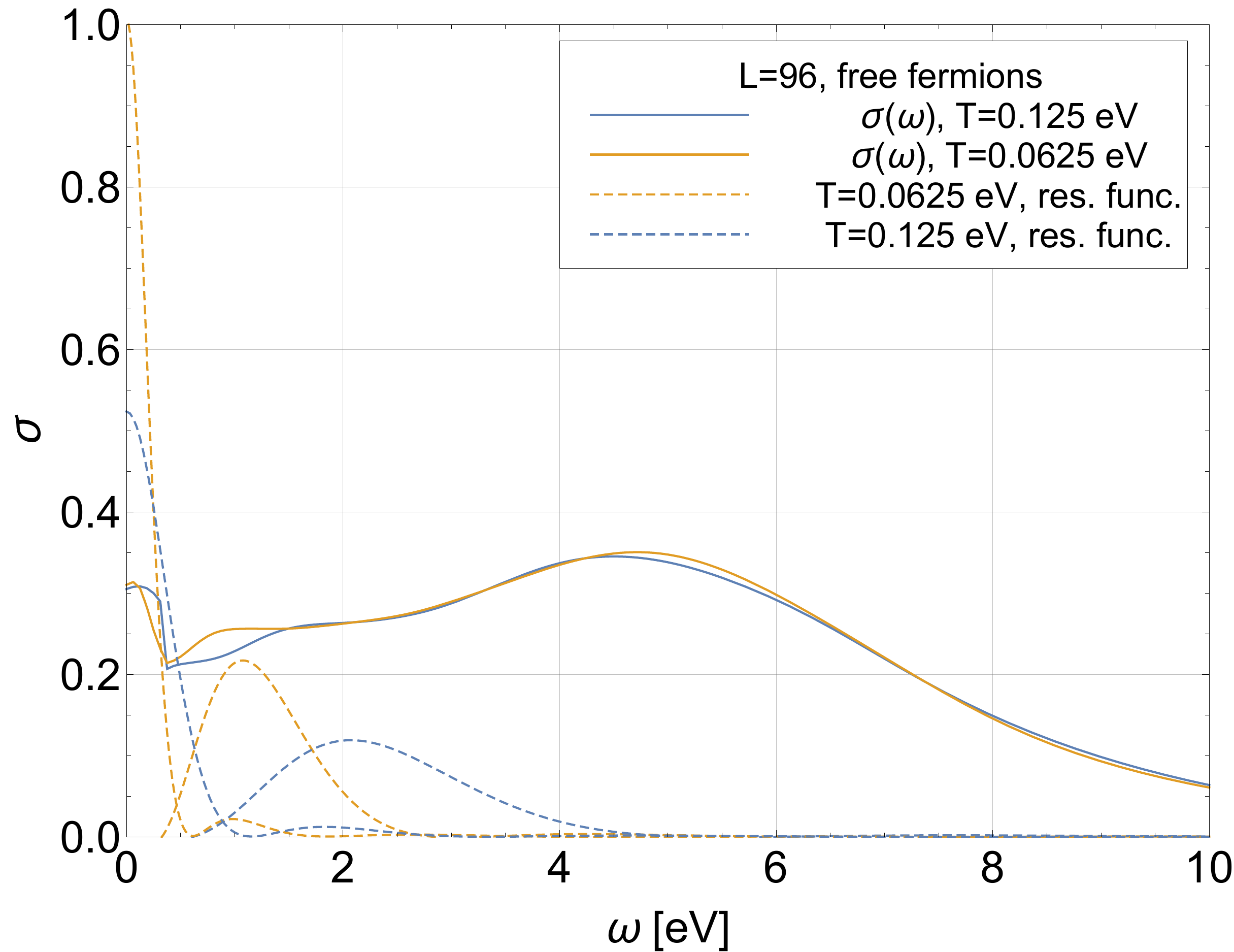}	
	\caption{Left: Calculation of the conductivity $\sigma(\omega)$ using the BG method from real QMC data obtained on a $48  \times 48$ lattice for graphene on a substrate with dielectric permittivity $\varepsilon=2$. The filled area shows the statistical error.  Right: Comparison of two BG reconstructions of the graphene conductivity at two temperatures. Here we used the exact current-current propagator for free fermions on a $96 \times 96$ lattice. We plotted both the spectral function and resolution functions at $\omega_0=0$ and near the plateau. }
	\label{fig:BG_check}
\end{figure}

The results of this analysis are shown in Fig.~\ref{fig:conductivity}.  One can see that only the BG method consistently reproduces all distinct features of the spectral function: the Drude peak at zero frequency, the drop-down at small $\omega$, the plateau at low energies and finally the peak at larger energies, $\omega\approx 5$~eV. The plateau also has the correct height, $\sigma \approx 0.25$. The Maximum Entropy Method shows a significant increase of the electrical conductivity at very small energies and therefore seems to be very sensitive to the Drude peak while a plateau at small energies is not visible. The SP method on the other hand shows neither a peak nor a plateau at small energies. Both the MEM as well as the SP method are however able to reconstruct the peak at $\omega\approx 5$~eV while the peak from the BG method is artificially broadened.  

The importance of the Drude peak for the reproduction of the initial Euclidean correlator data is demonstrated in the left panel of Fig.~\ref{fig:conductivity}. One can see that, since BG and MEM reproduce the Drude peak, their curves are the closest to the initial lattice data near the middle point of the correlator. In contrast, the SP method does not reproduce the peak and shows the largest discrepancy with respect to the Euclidean correlator data.

In general, the BG method reproduces the plateau accurately once the typical width of the resolution functions is smaller than the width of the plateau. Here we partially reproduce this analysis, following two previous papers on the subject: \cite{conductivity2} and  \cite{conductivity0}.  On the right panel of Fig.~\ref{fig:BG_check} we show the comparison of the BG reconstructions made at two different temperatures. The width of the resolution function  (at least at relatively low $\omega_0$) is proportional to the temperature, thus the quality of the reconstruction increases and the plateau feature becomes more prominent once the temperature decreases.  We also see how small the width of the resolution function is at  $\omega_0=0$: it guarantees the reconstruction of the Drude peak. It also explains the difference to Fig.~\ref{fig:lattice_data}, where the sharp peak and the step in the spectral function were not reproduced that well by the BG algorithm. The difference is that in the case of the graphene conductivity all these sharp features are located at low frequencies, where the resolution functions are substantially narrower.

It is also important to note that all these properties of the algorithm are not lost due to the errors in the real QMC data, which is demonstrated in Fig.~\ref{fig:BG_check} on the left panel: we are still able to reconstruct the plateau and the Drude peak, despite the appearance of some statistical uncertainties.

\section{Summary}
\label{sec:summary}

In this work we have performed a direct comparison of three different numerical analytic continuation methods each of which have their own merits: the Maximum Entropy Method, the Backus-Gilbert method and the Schlessinger point or Resonances Via Pad\'{e} method. In particular MEM and the Backus-Gilbert method are frequently used in order to extract spectral functions or other real-time properties from Euclidean input data. The SP (or RVP) method on the other hand has only rarely been used in this context. This work represents the first application of this method to data from lattice QCD and the other examples studied here.

First, we have applied these three methods to mock data obtained from a Breit-Wigner model spectral function and compared the reconstructions. The regime of applicability of these methods in dependence of the number of input points on the Euclidean correlator and their statistical error has been discussed. It was found that the SP method gives almost exact reconstructions if the number of input points is large enough and their error is small while MEM and the BG method also give reasonable reconstructions for larger errors, at least for positions of the peaks in the spectrum. For other features of the spectral function, sometimes the MEM, sometimes the BG method performs better, as summarized below.

We applied these reconstruction methods to Euclidean data on propagators obtained from a Functional Renormalization Group calculation for different temperatures in order to extract the corresponding spectral functions. This recently developed analytically continued aFRG setup allows to not only calculate the Euclidean propagators but also the real-time spectral functions directly. We were therefore able to compare the aFRG spectral functions with the ones obtained from the numerical continuation methods using the Euclidean propagator as input, and to use the aFRG spectral functions in the Lehmann representation in order to check the consistency with the Euclidean propagator for the first time.

Although the error of these data is very small and many data points are available the quality of the numerical reconstruction decreases with increasing temperature for all three methods, as expected. We found that the SP method always gives the best reconstruction for these data and is able to reconstruct the peak as well as the continuum of the spectral function rather well. MEM is also able to capture the dominant peaks but has difficulties with the continuum while the BG method yields only very rough estimates of the overall spectral functions.

When applied to lattice QCD data on the Euclidean correlator of the iso-vector vector current, we find that MEM performs better than the other two methods and is able to identify the main peak correctly. The BG method as well as the SP method are not able to reconstruct the spectral function which in the case of the SP method is clearly due to the large errors of the data points.

As a last example, we studied data obtained from a free tight-binding model for graphene in order to extract its electrical conductivity. In particular the plateau structure at low energies and hence the DC conductivity is best reconstructed by the BG method, which can be checked by switching to lower temperatures and/or better quality data, when we can reduce the width of the resolution functions.  On the other hand, the dominant peak at larger energies is more clearly seen by the MEM and the SP method.

In summary, the SP method has proven to be a potentially very powerful numerical method for analytic continuation that suggests itself as a serious alternative to the more well-established MEM and the BG method, if the errors of the input data are small enough and if the sampling rate is sufficiently high. This model-independent method might therefore be the preferred choice when applied to data from numerical solutions to functional equations such as the FRG flow equations for QFT correlation functions or also their Dyson-Schwinger equations \cite{Alkofer2001}.

For data with larger statistical errors and somewhat coarser sampling, as
typical for present-day, even large-scale lattice simulations, MEM is once again
confirmed as the most reliable method available to date.
Specifically, we observe that its particular strength is in the reconstruction of pronounced peak structures where MEM usually yields the most accurate estimates of both, the peak position and its width. For such sharp structures, that arise for example from narrow resonances or well separated states in a discrete spectrum, the BG method will most likely still give the correct estimate of the peak position, but the peak width is usually reconstructed not so accurately, because of the additional broadening that arises due to the finite widths of the resolution functions.

In contrast, the BG method is much more reliable in the reconstruction of some smooth features of spectral functions, like plateaus. For those to be reconstructed correctly, one only needs to reduce the width of the corresponding resolution function to be smaller then the width of the plateau. Once this requirement is satisfied, the BG method is very robust and provides the most stable results. The interpretation of the data and the error estimates are relatively straightforward with the BG method, while this is aggravated by the typical ringing that occurs in MEM which can produce unphysical fluctuations around the plateau and often makes it difficult to even recognize its position. Although MEM is more suitable for resonances, the more natural application of the BG method should therefore be the calculation of transport coefficients, for example. 

\section*{Acknowledgments}

The research of P.G. is supported 
by the Mext-Supported Program for the Strategic 
Foundation at Private Universities, ``Topological Science" (No. S1511006). R.-A.T. and L.v.S. acknowledge support by the Deutsche Forschungsgemeinschaft (DFG) through the grant CRC-TR 211 ``Strong-interaction matter under extreme conditions.'' M.U.~and L.v.S.~were also supported by the DFG grants BU 2626/2-1 and SM 70/3-1.    


\bibliography{qcd} 

\end{document}